\documentclass{iucrjournals}

\title{Theoretical Analysis of Topotomography Using Small Intragranular Strain Approximations}

\author[a,b]{Zheheng Liu\IUCrCemaillink{zheheng.liu@esrf.fr}}
\author[c]{Nicola Vigano\IUCrCemaillink{nicola.vigano@cea.fr}}
\author[d]{Henry Proudhon}
\author[a,b]{Wolfgang Ludwig\IUCrCemaillink{wolfgang.ludwig@esrf.fr}}

\affil[a]{MATEIS, INSA Lyon, UMR5510 CNRS, 69621 Villeurbanne, France}
\affil[b]{ESRF, Grenoble, France}
\affil[c]{Université Grenoble Alpes, CEA, IRIG-MEM, Grenoble, 38000, France}
\affil[d]{MINES Paris, PSL University, MAT - Centre des Matériaux, CNRS UMR 7633, BP 87, Evry, 91003, France}
\usepackage{amsmath}
\usepackage{graphicx}

\usepackage{float}
\usepackage{subfigure}

\usepackage{caption}
\begin{document} 
\maketitle
\makeatletter
\makeatother
\nolinenumbers

\begin{synopsis}
Theoretical analysis of the possibility to reconstruct the intragranular crystal orientation fields from acquisitions of near-field diffraction data, acquired during rotation of a grain around one of its lattice plane normals.
\end{synopsis}

\begin{abstract}
Topo-Tomography (TT) is a synchrotron-based X-ray diffraction imaging technique used to characterize grain shape and crystal orientation in polycrystalline samples. This work aims to provide a decisive and fundamental understanding of 3D grain shape and orientation field reconstruction from TT diffraction data. We derive mathematical expressions for the TT projection geometry, considering grain shape, intragranular lattice rotations, and elastic strains, under the assumption of kinematical diffraction. These expressions are simplified using approximations for small strain variations and grain size. The simplified expressions show that integrated TT projection images correspond to projections of a "pseudo" distorted grain volume. Its Fourier analysis provides insights into the feasibility of orientation field reconstruction from TT scans. We propose methods to expand data coverage, including using opposite scattering vectors and varying detector distance. A lower bound for orientation sampling resolution is derived and validated through simulations.
\end{abstract}

%
%
%
%
%


\section{Introduction}
Understanding the relationship between a material’s 3D crystalline microstructure and its macroscopic mechanical properties is essential to predict and prevent mechanical failure. There is consequently an increasing demand for non-destructive, 3D characterization of the crystalline microstructure in structural materials like metals and their alloys. Using Bragg diffraction and high-energy monochromatic X-ray beams at synchrotrons, X-ray diffraction imaging techniques can observe the 3D grain microstructure and crystal orientation fields in the bulk of polycrystalline samples. They offer spatial resolutions in the order of the micrometer, and angular resolutions in the order of a few hundredths of a degree. These techniques encompass a range of methods, each offering unique capabilities for visualizing and analyzing materials at various scales and levels of detail.


Among these techniques, Diffraction Contrast Tomography (DCT)~\cite{ludwig2009three} and Topo-Tomography (TT)~\cite{ludwig2001three, ludwig2007high} are notable for their ability to achieve fast scanning of 3D sample and grain volumes as they are both based on extended (box) beam illumination geometries. Unlike forward-modeling-based grain-mapping approaches~\cite{Suter2006, Li2013, Nygren2020}, DCT and TT take into account diffracted intensities and make use of reconstruction and optimization algorithms deployed in the field of tomographic imaging. With DCT one can reconstruct the spatial positions and lattice orientations of grains in a sample volume, providing a comprehensive overview of the microstructure in the polycrystalline material. TT is designed to scan individual grains within a larger sample volume, typically deploying a high-resolution configuration of the X-ray detector system. 

As outlined in~\cite{vigano2020x}, since the 2D projection images (i.e. the spatially resolved diffraction spots) of a given grain are affected by both the grain shape and the intragranular orientation field, the projection geometry is not a priori known and a 6D reconstruction framework~\cite{Vigano2014} can be used to jointly reconstruct the grain shape and the intragranular lattice orientations from DCT or TT scanning data. The 6D representation refers to a discrete sampling of 3D position and 3D orientation space~\cite{Vigano2014}, which linearizes the inverse problem and allows deploying iterative reconstruction algorithms used in the field of convex optimization~\cite{Sidky2012}. However, TT reconstructions may exhibit some artifacts that reduce the accuracy of the analysis. To address these challenges, a deeper understanding of the TT scan is desired to provide a theoretical basis for improvements of the scanning process and optimization of the parameters, ultimately enhancing the quality and reliability of TT reconstructions. Although experimental advances in TT are significant~\cite{Proudhon2018,Stinville2022}, a comprehensive theoretical framework to analyze and optimize the technique is still lacking.

In this study, we derive a mathematical expression for Topo-Tomography to establish a theoretical foundation for understanding the technique. The expression is then simplified using approximations based on small intragranular orientation spreads and grain size to make the analysis more practical. Fourier analysis has been performed using these simplified expressions. The joint use of opposite scattering vectors and/or a series of detector distances is proposed to increase the data coverage. Moreover, the analysis allows us to derive a lower bound for the orientation sampling resolution to be deployed in the reconstruction algorithm. These developments aim to reduce artifacts and improve the accuracy of TT grain reconstructions.

\subsection{Introduction of Topo-Tomography}

\begin{figure}[H]
    \centering
    \includegraphics[width=36pc]{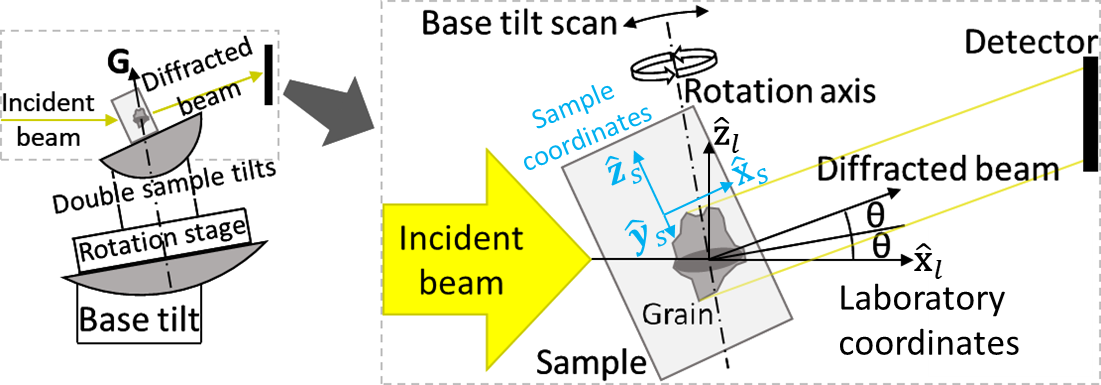}
    \caption{Sketch of Topo-Tomography (TT) setup and associated coordinates.}
    \label{fig:TTsketch}
\end{figure}

To implement the TT scan, the average crystal orientation of the target grain is required as input information. Typically, the lattice orientations and positions of all grains are obtained via the acquisition and reconstruction of a DCT scan prior to the TT experiment. According to the lattice orientation of the target grain, the sample is tilted by two goniometer tilt stages to align a selected scattering vector $\mathbf{G}$ of the target grain with the rotation axis. Depending on the grain position, the sample is shifted to place the target grain on the rotation axis. Finally, the rotation axis is tilted by the Bragg angle to satisfy the Bragg condition for the chosen crystal plane, as illustrated in figure~\ref{fig:TTsketch}. Then, the sample is rotated in steps about the rotation axis ($\omega$ angle; typically the outer scan loop), and the detector collects a series of diffraction blobs (3D image stacks) as the target grain is continuously scanned over a limited range of the base tilt rotation angle ($\phi$, typically inner scan loop). This base tilt scan around the $\hat{\mathbf{y}}_l$-axis covers the intragranular orientation spread of the target grain and allows rotating misoriented grain sub-volumes into diffraction condition. Therefore, TT scan data have four dimensions, which correspond to the two dimensions of the detector image, one dimension of sample rotation, and one dimension for the base tilt scan. Finally, the 3D grain morphology~\cite{ludwig2007high} and, more recently, also information on intragranular orientations can be reconstructed from these diffraction spots, using the 6D reconstruction framework described in~\cite{vigano2020x}.

Since only one crystal plane is used for a TT scan, the lattice rotations within the plane cannot be reconstructed, but only the normal vectors of the crystal plane can be reconstructed. The normal vectors can be described in a 2D lattice orientation space, so the grain reconstruction from a TT scan can be described in a 5D position-orientation space and the third orientation component along the scattering vector is kept constant. As outlined in~\cite{vigano2020x}, this 5D space can be viewed either as a collection of 2D orientation spaces (one for each real space voxel) or, alternatively, as a collection of real space volumes, each representing a slightly different component of the 2D orientation space sampling. For technical reasons, the latter representation has been adopted in the 6D reconstruction framework, taking advantage of GPU-accelerated forward and backprojection operations implemented in the ASTRA toolbox~\cite{vanAarle2016}. Note that at the end of the reconstruction process, the average of the 2D orientation sub-space is calculated for each of the real space voxels and only this average quantity is considered when evaluating the reconstruction results against the known ground truth in our simulations.

\subsection{Relationship between the scanning data and the grain}

For TT reconstruction, the 6D algorithm~\cite{vigano2020x} describes the grain in a discretized 5D position-orientation space, because the diffraction spots can be seen as linear projections of the 5D grain according to the Born approximation. According to~\cite{vigano2020x}, the reconstruction problem can be expressed as the optimization problem~\eqref{eq:6DDCT}.
\begin{equation}
    \label{eq:6DDCT}
    \mathbf{x}^* = \arg \min_{\mathbf{x}} 0.5\lVert\mathbf{y}-\mathbf{A}\mathbf{x}\rVert_2^2 + \lambda\lVert\mathbf{H}(\mathbf{x})\rVert_1\texttt{, subject~to~}\mathbf{x}\succeq\mathbf{0}.
\end{equation}
Here, bold lowercase letters represent column vectors, while bold uppercase letters denote matrices. The column vector $\mathbf{x}$ contains the scattering intensities of all the 5D samplings. $\mathbf{y}$ contains the intensities of all the diffraction spot pixels. $\mathbf{A}$ is the system matrix that describes the linear relationship between the scattering intensities distributed in the 5D position orientation space and the diffracted beam intensities received by the detector pixels. As $\mathbf{A}$ is typically ill-posed, the regularization term $\lambda\lVert\mathbf{H}(\cdot)\rVert_1$ is used. This term injects prior knowledge into the reconstruction and it has two-fold effect: it helps reducing the influence of noise and it mitigates the under-determinacy of the said ill-posed inverse problem. In particular, this is achieved thanks to $\mathbf{H}$, which is a suitable transformation that provides a parsimonious representation of the expected signal, and the $l_1$-norm, which imposes sparsity in the decomposition space of the reconstructed volume.

In~\cite{vigano2020x}, the linear relationship $\mathbf{y}$=$\mathbf{A}\mathbf{x}$ between the diffraction spots and the 5D grain is explained in detail using four coordinate systems and their transforms, which are the laboratory coordinates, the sample coordinates, the reconstruction coordinates and the detector coordinates.

As illustrated in figure \ref{fig:TTsketch}, the laboratory coordinates are represented by [$\hat{\mathbf{x}}_l$, $\hat{\mathbf{y}}_l$, $\hat{\mathbf{z}}_l$]. $\hat{\mathbf{x}}_l$ aligns with the incident beam. $\hat{\mathbf{z}}_l$ is perpendicular to $\hat{\mathbf{x}}_l$ and lies in the plane spanned by $\hat{\mathbf{x}}_l$ and the diffractometer $\omega$  rotation axis. $\hat{\mathbf{y}}_l$ is perpendicular to both $\hat{\mathbf{x}}_l$ and $\hat{\mathbf{z}}_l$, completing the right-handed coordinate system. 

The sample coordinates are represented by [$\hat{\mathbf{x}}_s$, $\hat{\mathbf{y}}_s$, $\hat{\mathbf{z}}_s$], which vary dynamically with the tilts and shifts of the sample goniometer stage and the diffractometer rotation and base tilt angles. The transform between the sample coordinates and the laboratory coordinates reflects the fixed settings of the sample goniometer and the motion of the diffractometer rotation and base tilt axes during TT scan acquisition. Typically, when there is no rotation, shift, or tilt, the sample coordinates coincide with the laboratory coordinates. The reconstruction coordinates define the arrangement of a discrete sampling grid used throughout the reconstruction process. A typical choice would be a shifted sample coordinate system with origin in the grain center position~\cite{vigano2020x}. To simplify the expressions, we will adopt a different setting introduced in section  section~\ref{sec:simpgeo}. 
The detector coordinates describe the spatial positions of the detector pixels in order to establish the mappings between the sample and the scanning data.

\section{Theory and Mathematical Foundation}

For theoretical analysis, the continuous grain and the continuous diffraction spot image are utilized to derive the general properties of the linear relationship between the grain and the scanning data of Topo-Tomography (TT). In~\cite{vigano2020x}, an integral form is derived to express this relationship, but the expression is mathematically complex and not straightforward to analyze. Therefore, a simplified expression is sought to facilitate a clearer and more manageable theoretical analysis, which would help to understand and optimize the TT scanning process. The simplifications must maintain the fundamental properties of the TT scan.

To simplify the mathematical expression of Topo-Tomography, an ideal and special setup of Topo-Tomography is first used to derive the expression of a 3D TT scan for an undeformed grain. The expression is then extended to the 5D TT case by considering variations in the lattice plane normals within deformed grains, followed by a further expansion to a 6D TT scan to involve intragranular elastic strain variations. For both the 5D and the 6D cases, approximations are introduced based on the assumptions of small intragranular strain variations and small grain size, which effectively simplify the resulting mathematical expressions. The simplified expressions for 5D and 6D TT scans are subsequently theoretically analyzed for deeper insight on the properties and information which can be extracted from TT scans.

\subsection{A special case to simplify the expression}\label{sec:simpgeo}

\begin{figure}[H]
    \centering
    \subfigure[Simplified reconstruction coordinates]{\label{fig:SketchSimpRecgeo}\includegraphics[width=15.11pc]{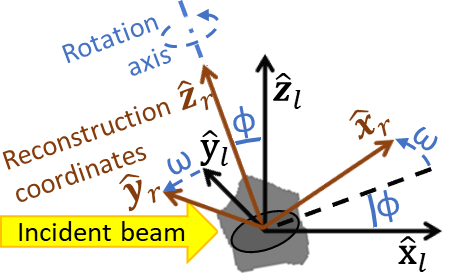}}
    \hspace{2pc}
    \subfigure[Reference coordinates]{\label{fig:SketchSimpRefdetgeo}\includegraphics[width=12.35pc]{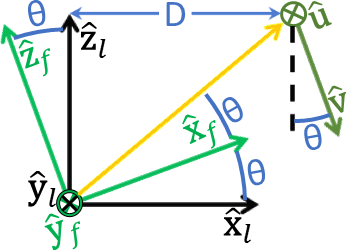}}
    \caption{Sketches of simplified coordinates used for formulation: (a) reconstruction coordinates used for formulation are equivalent to the rotation of laboratory coordinates based on $\phi$ and $\omega$; (b) reference coordinates used for formulation are equivalent to the rotation of laboratory coordinates around base tilt by -$\theta$.}
    \label{fig:SketchSimpGeos}
\end{figure}

An ideal and special case is introduced to simplify the coordinate transforms in the expression of TT scans. Let [$\hat{\mathbf{x}}_r$, $\hat{\mathbf{y}}_r$, $\hat{\mathbf{z}}_r$] indicate the three axes of the reconstruction coordinates. In this special case, the origin and the $\hat{\mathbf{z}}_r$ axis of the reconstruction coordinates are aligned with the rotation center and rotation axis of the sample rotation stage of the diffractometer. For the base tilt, its rotation center is also located at the origin and its rotation direction is aligned with the $\hat{\mathbf{y}}_l$ axis of the laboratory coordinates. Let $\omega$ and $\phi$ indicate the sample rotation angle and the base tilt rotation angle. As illustrated in figure~\ref{fig:SketchSimpRecgeo}, at the base tilt angle of $\phi$=0 and the rotation angle of $\omega$=0, the reconstruction coordinates [$\hat{\mathbf{x}}_r$, $\hat{\mathbf{y}}_r$, $\hat{\mathbf{z}}_r$] coincide with the laboratory coordinates [$\hat{\mathbf{x}}_l$, $\hat{\mathbf{y}}_l$, $\hat{\mathbf{z}}_l$].

As illustrated in figure~\ref{fig:SketchSimpRecgeo}, the reconstruction coordinates rotate as the base tilt angle $\phi$ or the rotation angle $\omega$ varies, which complicates the formulation, so a reference coordinate system represented by [$\hat{\mathbf{x}}_f$, $\hat{\mathbf{y}}_f$, $\hat{\mathbf{z}}_f$] is introduced to simplify the derivation of the mathematical expression, as illustrated in figure~\ref{fig:SketchSimpRefdetgeo}. Let $\theta$ indicate the Bragg angle of the lattice plane used for TT scan. The reference coordinates can be obtained from laboratory coordinates by rotating -$\theta$ about $\hat{\mathbf{y}}_l$. The reference coordinates do not change as the base tilt angle $\phi$ or the rotation angle $\omega$ varies (instead, the direction of the incoming beam and detector position are updated accordingly). At the base tilt angle of $\phi$=-$\theta$ and rotation angle of $\omega$=0, the reconstruction coordinates coincide with the reference coordinates.

Let [$\hat{\mathbf{u}}$, $\hat{\mathbf{v}}$] indicate the two axes of the detector coordinates. Ideally, $\hat{\mathbf{u}}$ is parallel to $\hat{\mathbf{y}}_l$ of the laboratory coordinates. As illustrated in figure~\ref{fig:SketchSimpRefdetgeo}, to simplify the formulation, the detector is tilted by the Bragg angle $\theta$ so that the $\hat{\mathbf{v}}$ axis of the detector is parallel to the $\hat{\mathbf{z}}_f$ axis of the reference coordinates. Let $D$ indicate the distance between the origin of the reference coordinates and the detector along the incident beam, i.e., the $\hat{\mathbf{x}}_l$ direction of the laboratory coordinates. In this special case, the origin of the detector is assumed to be [$D$, 0, $D\tan2\theta$] in laboratory coordinates, so that a voxel, which is at the origin of the reference coordinates and has the normal of the plane aligned along the $\hat{\mathbf{z}}_f$ axis, diffracts the beam to the origin of the detector.

\subsection{3D TT scan for undeformed grain}

We consider the idealized case of an undeformed grain for which the diffraction spots are not distorted by intragranular orientation spread and can be regarded as geometrical projections of the grain volume (neglecting attenuation along the path of the incoming and diffracted beam). Therefore, in this case, the TT scan is equivalent to a laminography~\cite{gondrom1999x} scan of the grain volume. The expression of the 3D TT scan using the special case is derived as follows.

\begin{figure}[H]
    \centering
    \subfigure[Projection of a voxel at origin.]{\label{fig:sketchvoxeloriginproj3D}\includegraphics[width=12.82pc]{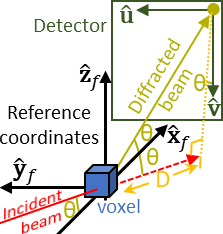}}
    \hspace{4.5pc}
    \subfigure[Projection of a voxel in $\hat{\mathbf{y}}_f$-$\hat{\mathbf{z}}_f$ plane.]{\label{fig:sketchvoxelshiftyzrefcoord}\includegraphics[width=15.52pc]{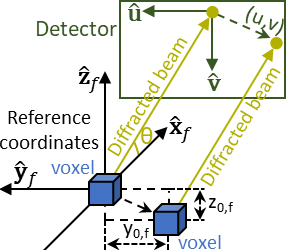}}
    \caption{Sketches depicting the projections of the voxels in the reference coordinates onto the detector coordinates. The $\hat{\mathbf{y}}_f$-$\hat{\mathbf{z}}_f$ plane is parallel to the plane of the detector.}
    \label{fig:sketchvoxel}
\end{figure}

Using the reference coordinates introduced in section  section~\ref{sec:simpgeo}, the origin of the detector coordinates is assumed to receive the diffracted beam from the voxel at the origin of the reference coordinates, as illustrated in figure~\ref{fig:sketchvoxeloriginproj3D}. The normal of the crystal plane is aligned with the $\hat{\mathbf{z}}_f$ axis. The direction of the diffracted beam is expressed as $\hat{\mathbf{d}}_f$=[$\cos\theta$, $0$, $\sin\theta$]$^T$ in the reference coordinates.

Let [$u$, $v$]$^T$ describe a point in the detector coordinates [$\hat{\mathbf{u}}$, $\hat{\mathbf{v}}$]. As introduced in  section~\ref{sec:simpgeo}, we suppose that the $\hat{\mathbf{u}}$ direction is the same as the $\hat{\mathbf{y}}_f$ direction of the reference coordinates and the $\hat{\mathbf{v}}$ direction is opposite to the $\hat{\mathbf{z}}_f$ direction. Let $\mathbf{r}_{0,f}$=[0, $y_{0,f}$, $z_{0,f}$]$^T$ indicate a point in the $\hat{\mathbf{y}}_f$-$\hat{\mathbf{z}}_f$ plane of the reference coordinates. As illustrated in figure~\ref{fig:sketchvoxelshiftyzrefcoord}, the detector coordinates of the projection of the voxel at $\mathbf{r}_{0,f}$ satisfy $u$=$y_{0,f}$ and $v$=$-z_{0,f}$. Let $\mathbf{r}_{f}$=[$x_f$, $y_f$, $z_f$]$^T$ indicate the reference coordinates. So, [u, v] receives the projections of the voxels on the line described by~\eqref{eq:3DTTlinefunc} in the reference coordinates. The operator $\times$ represents the cross product.
\begin{equation}
    \label{eq:3DTTlinefunc}
    \mathbf{0}=\hat{\mathbf{d}}_f\times(\mathbf{r}_f-\mathbf{r}_{0,f})= \cos\theta
    \begin{bmatrix}
        (u-y_f)\tan\theta & x_f\tan\theta-z_f-v & y_f-u \\
    \end{bmatrix}^T.
\end{equation}

Ideally, the base tilt scan is not required for a strictly undeformed grain, so the transform between the reconstruction coordinates and the reference coordinates can be expressed as~\eqref{eq:3DTTrefgeo2recgeoformulate}, where $\mathbf{r}_r$ represents a point in the reconstruction coordinates [$\hat{\mathbf{x}}_r$, $\hat{\mathbf{y}}_r$, $\hat{\mathbf{z}}_r$] and $\mathbf{\Omega}_\omega$ represent the rotation matrix of the sample rotation $\omega$ respectively.
\begin{equation}
    \label{eq:3DTTrefgeo2recgeoformulate}
    \mathbf{r}_{f}=
    \begin{bmatrix}
        x_f \\
        y_f \\
        z_f \\
    \end{bmatrix}=\mathbf{\Omega}_\omega\mathbf{r}_r=
    \begin{bmatrix}
        \cos\omega & -\sin\omega & 0 \\
        \sin\omega & \cos\omega & 0 \\
        0 & 0 & 1 \\
    \end{bmatrix}
    \begin{bmatrix}
        x \\
        y \\
        z \\
    \end{bmatrix}.
\end{equation}
The constraints can be applied using the delta function $\delta(\cdot)$. Therefore, the 3D TT scan for the undeformed grain can be expressed as~\eqref{eq:3DTTformulation} in the special case.
\begin{equation}
    \label{eq:3DTTformulation}
    \begin{split}
        B_{TT}(u,v,\omega) \propto \iiint_V & X_{3D}(\mathbf{r}_r)\delta(\mathbf{d}_f\times(\mathbf{\Omega}_\omega\mathbf{r}_r-\mathbf{r}_{0,f}))dV \\
        \propto \iiint_V & X_{3D}(x,y,z) \delta(z+v+(y\sin\omega-x\cos\omega)\tan\theta) \\
        & \delta(x\sin\omega+y\cos\omega-u)dxdydz.
    \end{split}
\end{equation}
Based on~\eqref{eq:3DTTformulation}, the Fourier transform of the diffraction spot $B_{TT}$ along $u$ and $v$ can be calculated as~\eqref{eq:3DTTformulationfourier}, which is denoted by $\hat{B}_{TT}(\cdot)$:
\begin{equation}
    \label{eq:3DTTformulationfourier}
    \begin{split}
        \hat{B}_{TT}(\rho_u,\rho_v,\omega)= & \iint B(u,v,\omega)e^{-j(\rho_uu+\rho_vv)}dudv \\
        \propto & \iiint_V X_{3D}(x,y,z)\exp\{j\rho_vz-j(\rho_u\sin\omega+\rho_v\tan\theta\cos\omega)x\\
        & \hspace{2pc} -j(\rho_u\cos\omega-\rho_v\tan\theta\sin\omega)y\}dxdydz.\\
    \end{split}
\end{equation}
The Fourier transform of the 3D grain volume $X_{3D}$ is instead expressed as~\eqref{eq:3DTTgrainvolfouriertransform}, which is denoted by $\hat{X}_{3D}$.
\begin{equation}
    \label{eq:3DTTgrainvolfouriertransform}
    \hat{X}_{3D}(\rho_x,\rho_y,\rho_z)=\iiint_V X_{3D}(x,y,z)\exp\{-j(\rho_xx+\rho_yy+\rho_zz)\}dxdydz.
\end{equation}
When comparing~\eqref{eq:3DTTformulationfourier} and~\eqref{eq:3DTTgrainvolfouriertransform}, we can see that $\hat{B}_{TT}(\cdot)$ contains the components of $\hat{X}_{3D}(\cdot)$, where the variables are transformed as~\eqref{eq:3DTTgrainvolfouriertransformvars}:
\begin{equation}
    \label{eq:3DTTgrainvolfouriertransformvars}
    \begin{bmatrix}
        \rho_x \\
        \rho_y \\
    \end{bmatrix}=
    \begin{bmatrix}
        \sin\omega & \cos\omega \\
        \cos\omega & -\sin\omega \\
    \end{bmatrix}
    \begin{bmatrix}
        \rho_u \\
        \rho_v\tan\theta \\
    \end{bmatrix};\,\rho_z = -\rho_v.
\end{equation}
The inequality $\rho_x^2+\rho_y^2\geq\rho_z^2\tan^2\theta$ can be derived from~\eqref{eq:3DTTgrainvolfouriertransformvars}, resulting in the Fourier double cone (FDC)~\cite{acciavatti2013oblique}. This means that a TT scan cannot obtain the Fourier components of the grain volume in the region of the two cones. As the frequencies in the cones are missing, the spatial resolution along $\hat{\mathbf{z}}_f$ is reduced. The lower the Bragg angle $\theta$, the smaller the region of the two cones, which improves the spatial resolution along $\hat{\mathbf{z}}_f$. This in turn suggests the use of low index $hkl$ reflections and use of elevated X-ray energies for optimal reconstruction results.

\subsection{5D TT with only variations in lattice plane normals}
\label{sec:formulaonlyrot}

At the base tilt angle of $\phi$=$-\theta$, the rotation axis and the average normal of the lattice plane are aligned along the $\hat{\mathbf{z}}_f$ axis in the reference coordinates. We usually let the base tilt angle $\phi$ scan around $-\theta$ so that the diffraction spot is steered upward, i.e. above the diffractometer, to avoid collisions with the detector system. Due to the spread of intragranular orientation, the plane can have local normals different from the $\hat{\mathbf{z}}_f$ direction. Let the unit vector $\hat{\mathbf{n}}_r$ = [$n_x$, $n_y$, $n_z$]$^T$ indicate the local normal vector of the lattice plane in the reconstruction coordinates. Let $\Delta\phi$ indicate the difference between the base tilt angle $\phi$ and the Bragg angle $\theta$, expressed as $\Delta\phi = \phi - (- \theta)=\phi+\theta$.

\begin{figure}[H]
    \centering
    \subfigure[The relationship of $\Delta\phi$, $\Delta\eta$ and plane normal.]{\label{fig:TTplnormphietasketch}\includegraphics[width=18 pc]{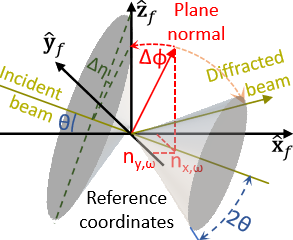}}
    \subfigure[Projection position changed by $\Delta\eta$.]{\label{fig:TTDetEtaSketch}\includegraphics[width=18 pc]{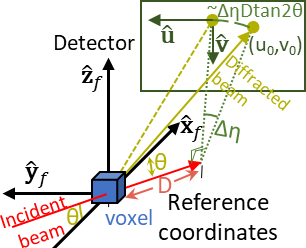}}
    \caption{Sketches depicting the diffraction condition and the projection position of a voxel which has undergone some lattice rotation.}
    \label{fig:TTdiffsketch}
\end{figure}

Figure~\ref{fig:TTplnormphietasketch} shows the sketch of $\Delta\phi$ required for a local plane normal to satisfy the Bragg condition. $n_{x,\omega}$ and $n_{y,\omega}$ are the first two elements of the rotated normal vector $\mathbf{\Omega}_\omega\hat{\mathbf{n}}_r$. When the normal of the lattice plane is located on the left cone, the Bragg condition is fulfilled and the diffracted beam will be on the right diffraction cone. The base tilt angle is changed by $\Delta\phi$ to align the normal with the cone. The position of the normal on the circle determines $\Delta\eta$.

Therefore, a diffraction condition can be expressed as $\mathbf{f}_{\phi\eta}$=$\mathbf{0}$, where the $\mathbf{f}_{\phi\eta}$ is the vector function defined in~~\eqref{eq:phietamatrixapproxfunc}. The detailed derivation is given in Appendix \ref{app:TTformulaplast}.
\begin{equation}
    \label{eq:phietamatrixapproxfunc}
    \begin{split}
        & \mathbf{f}_{\phi\eta}(\Delta\phi,\Delta\eta,n_x,n_y,\omega)=\\
        &
        \begin{bmatrix}
        n_x\cos\omega -n_y\sin\omega-\sin\Delta\phi+2\sin^2(\Delta\eta/2)\sin\phi\cos\theta \\
        n_x\sin\omega +n_y \cos\omega-\sin\Delta\eta\cos\theta \\
    \end{bmatrix}.\\
    \end{split}
\end{equation}

We suppose that for $\Delta\eta$=0 the beam diffracted from the origin of the reference coordinates will point to the origin of the detector. Contrarily, as illustrated in ~figure~\ref{fig:TTDetEtaSketch}, when $\Delta\eta$ is not 0, the diffracted beam will arrive at a point different from the origin on the detector. We denote this point as [$u_0$, $v_0$]$^T$, which is expressed as~\eqref{eq:TTdiffu0v0noderiv}. The detailed derivation is given in Appendix \ref{app:TTformulaplast}.
\begin{equation}
    \label{eq:TTdiffu0v0noderiv}
    \begin{bmatrix}
        u_0 \\
        v_0 \\
    \end{bmatrix}=\dfrac{D}{1-2\sin^2(\Delta\eta/2)\sin2\theta\tan\theta}
    \begin{bmatrix}
        -\sin\Delta\eta\tan2\theta \\
        4\sin^2(\Delta\eta/2)\sin\theta \\
    \end{bmatrix}.
\end{equation}
Let [$u$, $v$]$^T$ indicate the position on the detector which receives the diffracted beam from the position $\mathbf{r}_{0,f}$=[0, $y_{0,f}$, $z_{0,f}$]$^T$ in the reference coordinates. As illustrated in figure~\ref{fig:sketchvoxelshiftyzrefcoord},~\eqref{eq:refposdiffcoord} can be obtained. 
\begin{equation}
    \label{eq:refposdiffcoord}
    \mathbf{r}_{0,f}=
    \begin{bmatrix}
        0 & y_{0,f} & z_{0,f} \\
    \end{bmatrix}^T=
    \begin{bmatrix}
        0 & u-u_0 & -(v-v_0) \\
    \end{bmatrix}^T.
\end{equation}
Let $\hat{\mathbf{d}}_f$ denote the diffracted beam direction in the reference coordinates, which is expressed as~\eqref{eq:TTdiffdirdiffnoderiv}. The detailed derivation is provided in Appendix \ref{app:TTformulaplast}. Then, the line of the diffracted beam can be described by the equations~\eqref{eq:diffbeamindiffcoord} in the reference coordinates.
\begin{equation}
    \label{eq:TTdiffdirdiffnoderiv}
    \hat{\mathbf{d}}_f=
    \begin{bmatrix}
        \cos\theta &
        0 &
        \sin\theta \\
    \end{bmatrix}-
    \begin{bmatrix}
        \tan\dfrac{\Delta\eta}{2}\sin\theta &
        1 &
        \tan\dfrac{\Delta\eta}{2}\cos\theta \\
    \end{bmatrix}\sin\Delta\eta\sin2\theta.
\end{equation}
\begin{equation}
    \label{eq:diffbeamindiffcoord}
    \hat{\mathbf{d}}_f\times(\mathbf{r}_f-\mathbf{r}_{0,f})=\mathbf{0}.
\end{equation}
The reference coordinates can be transformed from the reconstruction coordinates according to the base tilt and the sample rotation as~\eqref{eq:possam2diff}, where $\mathbf{\Omega}_{\Delta\phi}$ represents the rotation matrix of the base tilt.

\begin{equation}
    \label{eq:possam2diff}
    \mathbf{r}_f=
    \mathbf{\Omega}_{\Delta\phi}\mathbf{\Omega}_\omega\mathbf{r}_r=
    \begin{bmatrix}
        \cos\Delta\phi & 0 & \sin\Delta\phi \\
        0 & 1 & 0 \\
        -\sin\Delta\phi & 0 & \cos\Delta\phi \\
    \end{bmatrix}
    \begin{bmatrix}
        \cos\omega & -\sin\omega & 0 \\
        \sin\omega & \cos\omega & 0 \\
        0 & 0 & 1 \\
    \end{bmatrix}
    \begin{bmatrix}
        x \\
        y \\
        z \\
    \end{bmatrix}.
\end{equation}
Let $X_{5D}$ and $B_{TT}$ indicate the 5D space of the grain and the 4D diffraction spot of the TT scan, respectively. Then, the TT scan can be expressed as~\eqref{eq:TTformula}, where $dO$=$dn_xdn_y$ and $dV$=$dxdydz$.
\begin{equation}
    \label{eq:TTformula}
    \begin{split}
        B_{TT}(u, v, \omega, \Delta\phi) = \iiint_V\iint_O & X_{5D}(x,y,z,n_x,n_y)\delta(\hat{\mathbf{d}}_f\times(\mathbf{\Omega}_{\Delta\phi}\mathbf{\Omega}_\omega\mathbf{r}_r-\mathbf{r}_{0,f})) \\
        & \delta(\mathbf{f}_{\phi\eta}(\Delta\phi,\Delta\eta,n_x,n_y,\omega)) dOdV.
    \end{split}
\end{equation}

\subsection{6D TT involving elastic deformation}\label{sec:formulaTTelastic}

Elastic strain can change the lattice spacing $d$. Along the normal of the crystal plane, the elastic strain $\epsilon_e$ equals $\Delta d$/$d$, where $\Delta d$ is the change in lattice spacing. According to Bragg's law $2d\sin\theta$=$\lambda$, the change in the lattice spacing can change the Bragg angle from $\theta$ to $\theta_e$ as~\eqref{eq:elasticstrainbragg}. Therefore, when considering elastic strain, the grain needs to be described in a 6D space for TT reconstruction, where 3 dimensions represent the position, 2 dimensions represent the normal vector of the plane and 1 dimension represents elastic strain.
\begin{equation}
    \label{eq:elasticstrainbragg}
    (1-\epsilon_e)\sin\theta_e=\sin\theta,\,\theta_e=\theta+\Delta\theta.
\end{equation}

Variations of the Bragg angle will result in changes in the required base tilt angle for diffraction, the diffraction direction, and the projection position on the detector. The diffraction condition involving $\Delta\phi$ and $\Delta\eta$ becomes $\mathbf{f}_{\phi\eta\Delta\theta}$=$\mathbf{0}$, where the vector function $\mathbf{f}_{\phi\eta\Delta\theta}$ is expressed as~\eqref{eq:phietamatrixapproxelasticfunc}. The detailed derivation is provided in Appendix \ref{app:TTformulaelast}.
\begin{equation}
    \label{eq:phietamatrixapproxelasticfunc}
    \begin{split}
        & \mathbf{f}_{\phi\eta\Delta\theta}(\Delta\phi,\Delta\eta,n_x,n_y,\omega,\Delta\theta) =\\
        &
        \begin{bmatrix}
        n_x\cos\omega-n_y\sin\omega+\sin(\Delta\phi+\Delta\theta)-(1-\cos\Delta\eta)\sin\phi\cos\theta_e \\
        n_x\sin\omega+n_y\cos\omega+\sin\Delta\eta\cos\theta_e \\
        \end{bmatrix}.
    \end{split}
\end{equation}
The updated projection positions $u_0$ and $v_0$, now including local variations of the lattice spacing, are expressed as~\eqref{eq:TTdiffu0v0elasticnoderiv}. The direction of the diffracted beam is expressed as~\eqref{eq:TTdiffdirdiffelastnoderiv} in the reference coordinates. The detailed derivation is provided in Appendix \ref{app:TTformulaelast}.
\begin{equation}
    \label{eq:TTdiffu0v0elasticnoderiv}
    \begin{bmatrix}
        u_0 &
        v_0 \\
    \end{bmatrix}^T=\dfrac{D
    \begin{bmatrix}
        -\sin\Delta\eta\sin2\theta_e\cos\theta &
        (1-\cos\Delta\eta)\sin2\theta_e\cos2\theta-\sin2\Delta\theta \\
    \end{bmatrix}^T}{(\cos(\theta+2\Delta\theta)-(1-\cos\Delta\eta)\sin2\theta_e\sin\theta)\cos2\theta}.
\end{equation}
\begin{equation}
    \label{eq:TTdiffdirdiffelastnoderiv}
    \hat{\mathbf{d}}_f=
    \begin{bmatrix}
        \cos(\theta+2\Delta\theta) \\
        0 \\
        \sin(\theta+2\Delta\theta) \\
    \end{bmatrix}-
    \begin{bmatrix}
        \tan(\Delta\eta/2)\sin\theta \\
        1 \\
        \tan(\Delta\eta/2)\cos\theta \\
    \end{bmatrix}\sin\Delta\eta\sin2\theta_e.
\end{equation}
Therefore, based on~\eqref{eq:TTformula}, when considering elastic strain, a TT scan can be expressed as~\eqref{eq:TTformulaelastic}, where $dO_e$=$dn_xdn_yd\epsilon_e$.
\begin{equation}
    \label{eq:TTformulaelastic}
    \begin{split}
        B_{TT}(u, v, \omega, \Delta\phi) = \iiint_V\iiint_{O_e} & X_{6D}(\mathbf{r}_r,n_x,n_y,\epsilon_e)\delta(\hat{\mathbf{d}}_f\times(\mathbf{\Omega}_{\Delta\phi}\mathbf{\Omega}_\omega\mathbf{r}_r-\mathbf{r}_{0,f}) \\
        & \delta(\mathbf{f}_{\phi\eta\Delta\theta}(\Delta\phi,\Delta\eta,n_x,n_y,\omega,\Delta\theta)) dO_edV.
    \end{split}
\end{equation}

\subsection{Formulation approximation}

\subsubsection{5D TT with only variations in lattice plane normals}

During early stages of plastic deformation, the intragranular misorientation angle is usually small and, hence, $\Delta\phi$ and $\Delta\eta$ are also small. As illustrated in figure~\ref{fig:TTplnormphietasketch}, since the rotation directions of $\Delta\phi$ and $\Delta\eta$ are the $\hat{\mathbf{y}}_l$ and $\hat{\mathbf{x}}_l$ directions in laboratory coordinates, $\Delta\phi$ and $\Delta\eta\cos\theta$ can be approximated as -$n_{x,\omega}$ and -$n_{y,\omega}$. Namely, according to the diffraction condition $\mathbf{f}_{\phi\eta}$=$\mathbf{0}$, $\Delta\phi$ and $\Delta\eta$ can be approximated as~\eqref{eq:approxphiplast}. The detailed derivation is provided in Appendix \ref{app:TTformulaplast}.
\begin{equation}
    \label{eq:approxphiplast}
    \Delta\phi \approx n_y\sin\omega-n_x\cos\omega;\,\Delta\eta \approx -(n_x\sin\omega+n_y\cos\omega)/\cos\theta.
\end{equation}
Since $\Delta\eta$ is small, based on~\eqref{eq:TTdiffu0v0noderiv}, $u_0$ and $v_0$ can be approximated as~\eqref{eq:smalletaondetapprox}, as illustrated in figure~\ref{fig:TTDetEtaSketch}. The approximation~\eqref{eq:smalletaondetapprox} can then be applied to~\eqref{eq:refposdiffcoord}.
\begin{equation}
    \label{eq:smalletaondetapprox}
    \begin{bmatrix}
        u_0 & v_0 \\
    \end{bmatrix}\approx
    \begin{bmatrix}
        -\Delta\eta D\tan2\theta & 0 \\
    \end{bmatrix}.
\end{equation}
Additionally, the grain size is assumed to be much smaller than the detector distance $D$. So, according to~\eqref{eq:TTdiffdirdiffnoderiv} and~\eqref{eq:diffbeamindiffcoord}, the line function of the diffracted beam within the grain can be approximated by~\eqref{eq:diffdirapprox}, which approximates the local direction of the diffracted beam as the average direction.
\begin{equation}
    \label{eq:diffdirapprox}
    \mathbf{0}=\hat{\mathbf{d}}_f\times(\mathbf{r}_f-\mathbf{r}_{0,f})\approx
    \begin{bmatrix}
        \cos\theta & 0 & \sin\theta \\
    \end{bmatrix}^T\times(\mathbf{r}_f-\mathbf{r}_{0,f}).
\end{equation}
Since $\Delta\phi$ is small,~\eqref{eq:possam2diff} can be approximated as~\eqref{eq:sam2diffcoordapproxphi}, which ignores the changes in the positions of the voxels caused by the variations $\Delta\phi$ in the base tilt angle.
\begin{equation}
    \label{eq:sam2diffcoordapproxphi}
    \mathbf{r}_f\approx
    \begin{bmatrix}
        x\cos\omega-y\sin\omega & x\sin\omega+y\cos\omega & z \\
    \end{bmatrix}^T.
\end{equation}

\begin{figure}[H]
    \centering
    \includegraphics[width=13pc]{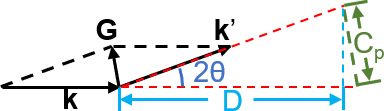}
    \caption{The sketch depicting the meaning of the coefficient $C_p$.}
    \label{fig:CpGeoMeaning}
\end{figure}

Therefore, the line function~\eqref{eq:diffbeamindiffcoord} can be further simplified by approximating~\eqref{eq:diffdirapprox} into~\eqref{eq:difflineapprox}, which describes the line of the diffracted beam within the grain. The factor $C_p$ is described in figure~\ref{fig:CpGeoMeaning}, where $\mathbf{G}$ represents the scattering vector and $\mathbf{k}$ and $\mathbf{k}'$ indicate the wave vectors of the incident beam and the diffracted beam, respectively.
\begin{equation}
    \label{eq:difflineapprox}
    \begin{bmatrix}
        (x+C_{p}n_x)\sin\omega+(y+C_{p}n_y)\cos\omega-u \\
        z+v+(y\sin\omega-x\cos\omega)\tan\theta \\
    \end{bmatrix}\approx\mathbf{0},\,C_{p}=D\tan2\theta/\cos\theta.
\end{equation}

As a result, the expression of the TT scan~\eqref{eq:TTformula} can be approximated to~\eqref{eq:TTformulaapprox}, which is simplified for theoretical analysis.
\begin{equation}
    \label{eq:TTformulaapprox}
    \begin{split}
        B_{TT}(u, v, \omega, \Delta\phi) \approx & \iiint_V\iint_O X_{5D}(x,y,z,n_x,n_y)\delta(\Delta\phi+n_x\cos\omega-n_y\sin\omega) \\
        & \delta((x+C_{p}n_x)\sin\omega+(y+C_{p}n_y)\cos\omega-u)dn_xdn_y \\
        & \delta(z+v+(y\sin\omega-x\cos\omega)\tan\theta)dxdydz. \\
    \end{split}
\end{equation}

\subsubsection{6D TT with elastic deformation}

When considering elastic strain, variations of Bragg angle will cause unignorable changes in the base tilt angle required for diffraction and the projection position on the detector in the $\hat{\mathbf{v}}$ direction. Namely, $\Delta\phi$ in~\eqref{eq:approxphiplast} and $v_0$ in~\eqref{eq:smalletaondetapprox} need to be expressed as~\eqref{eq:approxphielast}, where $\epsilon_e\approx\Delta\theta\cot\theta$.
\begin{equation}
    \label{eq:approxphielast}
    \Delta\phi \approx n_y\sin\omega-n_x\cos\omega-\Delta\theta;\,v_0 \approx -2D\Delta\theta/(\cos\theta\cos2\theta)\approx -C_p\epsilon_e/\cos^2\theta.
\end{equation}
Therefore, the expression of the TT scan in~\eqref{eq:TTformulaelastic} can be approximated by~\eqref{eq:TTformulaapproxelastic}.
\begin{equation}
    \label{eq:TTformulaapproxelastic}
    \begin{split}
        & B_{TT}(u, v, \omega, \Delta\phi) \approx \iiint_{O_e}\iiint_V \delta(\Delta\phi+\epsilon_e\tan\theta+n_x\cos\omega-n_y\sin\omega) \\
        & X_{6D}(x,y,z,n_x,n_y,\epsilon_e)\delta((x+C_{p}n_x)\sin\omega+(y+C_{p}n_y)\cos\omega-u) \\
        & \delta(z+C_p\epsilon_e/\cos^2\theta+v+(y\sin\omega-x\cos\omega)\tan\theta)dxdydzdn_xdn_yd\epsilon_e. \\
    \end{split}
\end{equation}

\subsection{3D reconstruction from integrated TT spots based on the 6D TT expression}\label{sec:3DTTrecdistortvol}

From the 6D TT expression~~\eqref{eq:TTformulaapproxelastic} simplified by the approximations, a remarkable property emerges: the integrated TT spots, using appropriate shifts, correspond to the 3D projections of a distorted grain volume. Below we provide the detailed derivation of this property, which is at the origin of apparently sharp, but geometrically distorted (!) grain volume reconstructions from integrated TT diffraction spots.

As illustrated in~\eqref{eq:TTformulaelasticintphi}, the TT spots are integrated over base tilt angle $\Delta\phi$, with the slices shifting $C_p\Delta\phi\tan\theta$ along the opposite of the $\hat{\mathbf{v}}$ direction.
\begin{equation}
    \label{eq:TTformulaelasticintphi}
    B_{TT,uv}(u, v, \omega) = \int B_{TT}(u, v+C_p\Delta\phi\tan\theta, \omega, \Delta\phi)d(\Delta\phi).
\end{equation}
Then, according to~\eqref{eq:TTformulaapproxelastic}, the integrated spots can be expressed as~\eqref{eq:TTformulaapproxelasticintphi}.
\begin{equation}
    \label{eq:TTformulaapproxelasticintphi}
    \begin{split}
        B_{TT,uv}(u, v, \omega) & \approx \iiint_O\iiint_V X_{6D}(x,y,z,n_x,n_y,\epsilon_e) \\
        & \delta(z+C_p\epsilon_e+v+((y+C_{p}n_y)\sin\omega-(x+C_{p}n_x)\cos\omega)\tan\theta) \\
        & \delta((x+C_{p}n_x)\sin\omega+(y+C_{p}n_y)\cos\omega-u)dxdydzdn_xdn_yd\epsilon_e.\\
    \end{split}
\end{equation}
Given the following equivalence:
\begin{equation}
    \label{eq:equicoord3D}
    \begin{bmatrix}
        x_{E} & y_{E} & z_{E} \\
    \end{bmatrix}=
    \begin{bmatrix}
        x & y & z \\
    \end{bmatrix}+C_p
    \begin{bmatrix}
        n_x & n_y & \epsilon_e \\
    \end{bmatrix}.
\end{equation}
The expression of the integrated TT spots can be rewritten from~\eqref{eq:TTformulaapproxelasticintphi} to~\eqref{eq:TTformulaapproxelasticintphi3D}, where the 6D grain $X_{6D}(x,y,z,n_x,n_y,\epsilon_e)$ is substituted with a 3D volume $X_{E}(x_{E},y_{E},z_{E})$ as~\eqref{eq:equigrainrec3D}.
\begin{equation}
    \label{eq:TTformulaapproxelasticintphi3D}
    \begin{split}
        B_{TT,uv}(u, v, \omega) & \approx \iiint X_{E}(x_{E},y_{E},z_{E})\delta(x_{E}\sin\omega+y_{E}\cos\omega-u) \\
        & \delta(z_{E}+v+(y_{E}\sin\omega-x_{E}\cos\omega)\tan\theta)dx_{E}dy_{E}dz_{E} \\
    \end{split}
\end{equation}
According to~\eqref{eq:equigrainrec3D}, $X_{E}(x_{E},y_{E},z_{E})$ can be regarded as a grain volume reconstruction distorted by intragranular orientation spread. According to~\eqref{eq:TTformulaapproxelasticintphi3D}, $B_{TT,uv}$ can be regarded as the geometric projections of the distorted grain volume $X_{E}$. This feature will be tested using the simulation data in  section~\ref{sec:grainphantomandsimspots}.
\begin{equation}
    \label{eq:equigrainrec3D}
    X_{E}(x_{E},y_{E},z_{E})=
    \iiint X_{6D}(
    x_{E}-C_pn_x,y_{E}-C_pn_y,z_{E}-C_p\epsilon_e,
    ,n_x,n_y,\epsilon_e)dO_e. 
\end{equation}

\subsection{Fourier analysis of 5D expression}

Fourier analysis is applied to the TT expression to gain deeper insights into the ability to reconstruct local lattice plane normal directions from TT projection data. Here, only the 5D case is considered to simplify the analysis. Inspired by the inverse Radon transform, the Fourier transform is performed to the 4D spot $B_{TT}$ along the dimensions of the detector image and base tilt ($u$, $v$, and $\Delta\phi$), as illustrated in~\eqref{eq:4DTTspotfft}.
\begin{equation}
    \label{eq:4DTTspotfft}
    \hat{B}_{TT}(\rho_u, \rho_v, \omega, \rho_{\Delta\phi})=\iiint B_{TT}(u, v, \omega, \Delta\phi) e^{-j(\rho_u u+\rho_v v+\rho_{\Delta\phi}\Delta\phi)}dudvd(\Delta\phi).
\end{equation}
~\eqref{eq:4DTTspotfft} can be calculated as~\eqref{eq:TTformulafft} based on the simplified expression~\eqref{eq:TTformulaapprox}. Furthermore, the 5D Fourier transform of the 5D grain can be expressed as~\eqref{eq:5Dgrainfft}. Comparing~\eqref{eq:TTformulafft} and~\eqref{eq:5Dgrainfft}, the Fourier transform of the 4D spot yields components of the Fourier transform of the 5D grain. This can be seen when using the variable transform ~\eqref{eq:fftvarsubs}.
\begin{equation}
    \label{eq:TTformulafft}
    \begin{split}
        & \hat{B}_{TT}(\rho_u, \rho_v, \omega, \rho_{\Delta\phi}) \approx \iint_O\iiint_V X_{5D}(x,y,z,n_x,n_y)\exp(j\rho_vz\\
        & -j(\rho_u\sin\omega+\rho_v\tan\theta\cos\omega)x-j(\rho_u\cos\omega-\rho_v\tan\theta\sin\omega)y\\
        & -j(C_p\rho_u\sin\omega-\rho_{\Delta\phi}\cos\omega)n_x-j(C_p\rho_u\cos\omega+\rho_{\Delta\phi}\sin\omega)n_y)dVdO. \\
    \end{split}
\end{equation}
\begin{equation}
    \label{eq:5Dgrainfft}
    \hat{X}_{5D}(\rho_x,\rho_y,\rho_z,\rho_{o,x},\rho_{o,y})=\iint_O\iiint_V X_{5D}e^{-j(\rho_xx+\rho_yy+\rho_zz+\rho_{o,x}n_x+\rho_{o,y}n_y)}dVdO.
\end{equation}
\begin{equation}
    \label{eq:fftvarsubs}
    \begin{bmatrix}
        \rho_x & \rho_{o,x}\\
        \rho_y & \rho_{o,y}\\
    \end{bmatrix}=
    \begin{bmatrix}
        \cos\omega & \sin\omega\\
        -\sin\omega & \cos\omega\\
    \end{bmatrix}
    \begin{bmatrix}
        \rho_v\tan\theta & -\rho_{\Delta\phi}\\
        \rho_u & C_p\rho_{u}\\
    \end{bmatrix};\,\rho_z=-\rho_v.
\end{equation}

\subsubsection{Increasing Fourier transform components}\label{sec:incrfftcomp}

To fully reconstruct the 5D grain, the full 5D domain ($\rho_x$,$\rho_y$,$\rho_z$,$\rho_{o,x}$,$\rho_{o,y}$) of the Fourier transform of the 5D grain is required. However,~\eqref{eq:TTformulafft} can only yield part of the Fourier transform of the 5D grain. For the purpose of clarification, the Fourier transform variables are replaced by~\eqref{eq:fftvarsubsforsimp}.
\begin{equation}
    \label{eq:fftvarsubsforsimp}
    \begin{bmatrix}
        \rho_x \\
        \rho_y \\
        \rho_z\\
    \end{bmatrix}=\dfrac{\Tilde{\rho}_r}{\cos\Tilde{\alpha}}
    \begin{bmatrix}
        {\sin(\Tilde{\alpha}+\omega)} \\
        {\cos(\Tilde{\alpha}+\omega)} \\
        {\sin\Tilde{\alpha}}/{\tan\theta}\\
    \end{bmatrix};\,
    \begin{bmatrix}
        \rho_{o,x}\\
        \rho_{o,y}\\
    \end{bmatrix}=
    \begin{bmatrix}
        -\cos\omega & \sin\omega \\
        \sin\omega & \cos\omega \\
    \end{bmatrix}
    \begin{bmatrix}
        \Tilde{\rho}_{o,x}\\
        \Tilde{\rho}_{o,y}\\
    \end{bmatrix}.
\end{equation}

\begin{figure}[H]
    \centering
    \includegraphics[width=30pc]{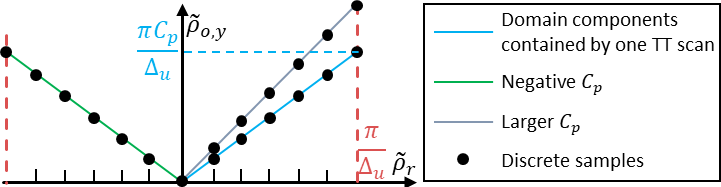}
    \caption{Sketch showing the components of $\Tilde{\rho}_r$-$\Tilde{\rho}_{o,y}$ domain contained by the TT scan data. Only quadrants I and II are shown, since quadrants IV and III are conjugate symmetric to I and II about the origin.}
    \label{fig:sketchFFTCompuCu}
\end{figure}

~\eqref{eq:fftvarsubsforsimpfinal} is derived according to~\eqref{eq:fftvarsubs} and~\eqref{eq:fftvarsubsforsimp}. As $\rho_u$, $\rho_v$, $\omega$, and $\rho_{\Delta\phi}$ are independent, a single TT scan can obtain the full domain of $\Tilde{\alpha}$, $\omega$ and $\Tilde{\rho}_{o,x}$ as well as either $\Tilde{\rho}_r$ or $\Tilde{\rho}_{o,y}$. However, $\Tilde{\rho}_r$ and $\Tilde{\rho}_{o,y}$ are proportional, so the TT scan data contain only a few components of the $\Tilde{\rho}_r$-$\Tilde{\rho}_{o,y}$ domain, as shown by the blue line in figure~\ref{fig:sketchFFTCompuCu}. Note that $\Tilde{\rho}_{o,y}$=$\rho_{o,y}\cos\omega$+$\rho_{o,x}\sin\omega$ and $\Tilde{\rho}_r^2$=$\rho_x^2+\rho_y^2-\rho_z^2\tan^2\theta$. Here, for a better understanding, a brief explanation of the variables $\Tilde{\rho}_r$, $\Tilde{\alpha}$ and $\omega$ is provided. A specific $\Tilde{\rho}_r$ determines a hyperboloid in the ($\rho_x,\rho_y,\rho_z$) domain, with the Fourier double cone serving as the asymptotic conical surface. The hyperboloid can be generated by rotating an inclined line around the $\rho_z$ axis. The angle between the inclined line and the $\rho_z$ axis is the Bragg angle $\theta$, and the distance between the inclined line and the origin is $\Tilde{\rho}_r$. $\omega$ is the rotation angle of the inclined line. $\Tilde{\alpha}$ is an angular coordinate based on $\omega$ to determine a point in the inclined line.
\begin{equation}
    \label{eq:fftvarsubsforsimpfinal}
    \begin{bmatrix}
        \Tilde{\rho}_r & \tan\alpha & \Tilde{\rho}_{o,x} & \Tilde{\rho}_{o,y} \\
    \end{bmatrix}=
    \begin{bmatrix}
        \rho_u & \rho_v\tan\theta/\rho_u & \rho_{\Delta\phi} & C_p\rho_u \\
    \end{bmatrix}.
\end{equation}

Since the Fourier transform of the real-valued data has conjugate symmetry about the origin, figure~\ref{fig:sketchFFTCompuCu} only shows quadrants I and II of the $\Tilde{\rho}_r$-$\Tilde{\rho}_{o,y}$ domain. A single TT scan contains the components on one single line of the $\Tilde{\rho}_r$-$\Tilde{\rho}_{o,y}$ domain.

To obtain the components in quadrants II and IV, a negative $C_p$ can be used, as shown by the green line in quadrant II of figure~\ref{fig:sketchFFTCompuCu}. Since $C_p$=$D\tan2\theta$/$
\cos\theta$ and the detector distance $D$ cannot be negative, the negative Bragg angle $\theta$ can be used, which represents the opposite scattering vector in the sample coordinates.

Using different detector distances $D$ can result in different values of $C_p$ to obtain more components of the $\Tilde{\rho}_r$-$\Tilde{\rho}_{o,y}$ domain. The two lines with different slopes in quadrant I of figure~\ref{fig:sketchFFTCompuCu} correspond to two TT scans using different detector distances $D$.

Combining TT scans with scanning of a vertical line beam across the grain (i.e. a 3D scanning procedure)  could be used to obtain more Fourier transform components. This geometry is analyzed in Appendix ~\ref{app:MoreTTScanMethods}.

\subsubsection{Orientation sampling resolution}
\label{sec:orisamresformula}

As the detector images and grain reconstruction are discretized, interpolation is required for the diffraction spots of the grain reconstruction. Linear interpolation is used here. The linear interpolation function and its Fourier transform are indicated by $f_{lin}(\cdot)$ and $\hat{f}_{lin}(\cdot)$, expressed as~\eqref{eq:linearinterpfunc}.
\begin{equation}
    \label{eq:linearinterpfunc}
    f_{lin}(x)=
    \begin{cases}
        1-\lvert x\rvert, & x\in(-1,1) \\
        0, & \text{otherwise} \\
    \end{cases};\,
    \hat{f}_{lin}(\rho)=\int f_{lin}(x)e^{-j\rho x}dx=\text{sinc}^2\dfrac{\rho}{2}.
\end{equation}
Let $X_{5D}^*$ and $B_{TT}^*$ indicate the grain reconstruction and its diffraction spots with linear interpolation. Based on~\eqref{eq:TTformulaapprox}, $B_{TT}^*$ can be expressed as~\eqref{eq:TTformulaapproxinterp}, where the operator $*$ indicates convolution. $\Delta_u$ and $\Delta_v$ represent the pixel sizes along $u$ and $v$. $\Delta_\phi$ represents the sampling interval of the base tilt scan. As the convolution leads to the product of the Fourier transforms, the Fourier transform of $B_{TT}^*$, denoted by $\hat{B}_{TT}^*$, can be expressed as~\eqref{eq:TTformulafftinterp}. The interpolations can be approximated as low-pass filters, leading to approximate cutoff angular frequencies of $2\pi\Delta_u^{-1}$, $2\pi\Delta_v^{-1}$ and $2\pi\Delta_\phi^{-1}$ for $\rho_u$, $\rho_v$ and $\rho_{\Delta\phi}$ of $\hat{B}_{TT}^*$ in~\eqref{eq:TTformulafftinterp}.
\begin{equation}
    \label{eq:TTformulaapproxinterp}
    \begin{split}
        & B_{TT}^*(u, v, \omega, \Delta\phi) \approx f_{lin}(\dfrac{u}{\Delta_u}) * f_{lin}(\dfrac{v}{\Delta_v}) * f_{lin}(\dfrac{\Delta\phi}{\Delta_\phi}) * \iiint_V\iint_O \\
        & X_{5D}^*(x,y,z,n_x,n_y) \delta((x+C_{p}n_x)\sin\omega+(y+C_{p}n_y)\cos\omega-u) \\
        & \delta(z+v+(y\sin\omega-x\cos\omega)\tan\theta)\delta(\Delta\phi+n_x\cos\omega-n_y\sin\omega)dOdV. \\
    \end{split}
\end{equation}
\begin{equation}
    \label{eq:TTformulafftinterp}
    \begin{split}
        & \hat{B}_{TT}^*(\rho_u, \rho_v, \omega, \rho_{\Delta\phi}) \approx \hat{f}_{lin}(\Delta_u\rho_u)\hat{f}_{lin}(\Delta_v\rho_v)\hat{f}_{lin}(\Delta_\phi\rho_{\Delta\phi})\iint_O\iiint_V \\
        & X_{5D}^*(\mathbf{r}_r,n_x,n_y) e^{j\rho_vz-j(\rho_u\sin\omega+\rho_v\tan\theta\cos\omega)x-j(\rho_u\cos\omega-\rho_v\tan\theta\sin\omega)y}\\
        & e^{-j(C_p\rho_u\sin\omega-\rho_{\Delta\phi}\cos\omega)n_x-j(C_p\rho_u\cos\omega+\rho_{\Delta\phi}\sin\omega)n_y}dVdO. \\
    \end{split}
\end{equation}

To avoid spectrum aliasing, orientation space sampling should meet the Nyquist rate. A sampling rate no less than twice the cutoff frequency can be used. Therefore, based on~\eqref{eq:fftvarsubs}, the sampling interval of $n_x$ should satisfy~\eqref{eq:noaliasingoresmaxfreq}, where $\Delta_o$ indicates the orientation sampling interval. The maximum in~\eqref{eq:noaliasingoresmaxfreq} can be obtained by [$C_p\Delta_u^{-1}$, $\Delta_\phi^{-1}$][$\sin\omega$, -$\cos\omega$]$^T\leq\rVert$[$C_p\Delta_u^{-1}$, $\Delta_\phi^{-1}$]$\lVert_2$. When $\tan\omega$=-$C_p\Delta_u^{-1}\Delta_\phi$, the maximum is reached. A similar inequality can be derived for the sampling interval of $n_y$.
\begin{equation}
    \label{eq:noaliasingoresmaxfreq}
    \Delta_{o}^{-1}\geq2\max_\omega(C_p\Delta_u^{-1}\sin\omega-\Delta_\phi^{-1}\cos\omega)=2\sqrt{C_p^2\Delta_u^{-2}+\Delta_\phi^{-2}}.
\end{equation}
For a given $\Delta_o$, in the 6D algorithm, spot data can be downsampled in the dimensions of the detector image and the base tilt ($u$, $v$, and $\Delta\phi$) to satisfy the inequality.





\section{Simulation results}

In this section, the simulation data are used to validate the three conclusions derived from the theoretical analysis of TT scans. These conclusions address the 3D reconstruction using integrated TT spots, the use of opposite scattering vectors, and the lower bound for orientation sampling resolution. A cubic grain phantom~\cite{joste2023simulation} with a slip band is used to generate simulated diffraction data and serves as the ground truth for comparison with the grain reconstructions. The phantom consists of 80$\times$80$\times$80 voxels with a voxel size of 0.37 $\mu m$ and features uniform scattering intensities throughout. Each voxel is assigned a lattice orientation. In the undeformed stage, all the phantom voxels have the same initial lattice orientation, which can be represented by the Rodrigues vector~\cite{frank1988orientation} of [-0.2857, 0.1429, 0.0001] in the fundamental zone~\cite{heinz1991representation} of the orientation space. As the phantom is deformed, each voxel has an intragranular misorientation (IGM) relative to the initial lattice orientation. These IGMs are characterized by 3D rotations of less than 0.025 degrees, which are too small to effectively test 5D TT reconstruction under practical conditions. Therefore, the original phantom is used only to test the conclusion regarding the 3D reconstruction using integrated TT spots, while the misorientation angles in the phantom are amplified by a factor of 10, which is used to test the other two conclusions.

\subsection{3D reconstruction from integrated TT spots}\label{sec:grainphantomandsimspots}

The simulation tool described in \cite{vigano2020x} is utilized to simulate diffraction blobs based on the original grain phantom. The energy of the X-ray beam is set to 50 keV and the detector pixel size of the TT spots is set to 1.22 $\mu m$. The simulated diffraction blobs are noiseless and intensity corrections like attenuation are ignored so that all simulated 3D blobs have identical total intensities.

\begin{figure}[H]
    \centering
    \subfigure[$D$=6 mm]{\label{fig:phantomttspotdist6}\includegraphics[width=12pc]{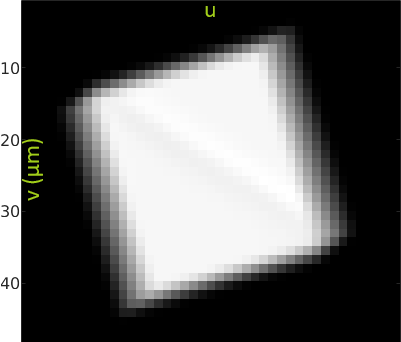}}
    \subfigure[$D$=20 mm]{\label{fig:phantomttspotdist20}\includegraphics[width=12pc]{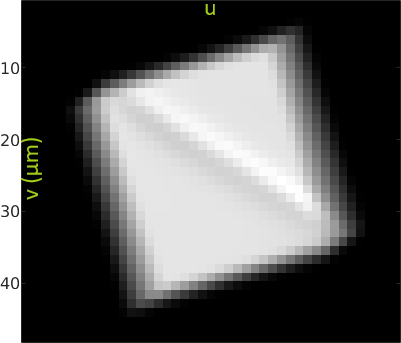}}
    \subfigure[$D$=50 mm]{\label{fig:phantomttspotdist50}\includegraphics[width=12pc]{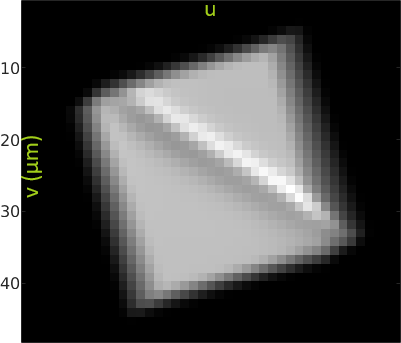}}
    \caption{The simulated diffraction spots of the phantom, integrated over base tilt angle $\phi$ at the same rotation angle $\omega$ with different detector distances $D$.}
    \label{fig:phantomttspotdist}
\end{figure}

Here, the (0 $\Bar{2}$ 2) crystal plane is chosen for the TT scan, resulting in a Bragg angle of $\theta$=5.687$^\circ$. Figure~\ref{fig:phantomttspotdist} shows the simulated diffraction spots of the original phantom, integrated over the base tilt angle $\phi$ at the same rotation angle $\omega$ with the detector distances $D$ of 6 mm, 20 mm and 50 mm. As the detector distance $D$ increases, the diffraction spot becomes increasingly distorted due to the intragranular orientation spreads. With a closer detector, the spots contain more information about the actual shape of the 3D volume; with a further detector, the spots contain more information about the variations in the normal vectors of the crystal plane within the grain.

\begin{figure}[H]
    \centering
    \subfigure[3D TT reconstruction]{\label{fig:3DTTrecvsdistortvol3DTTrec}\includegraphics[width=15pc]{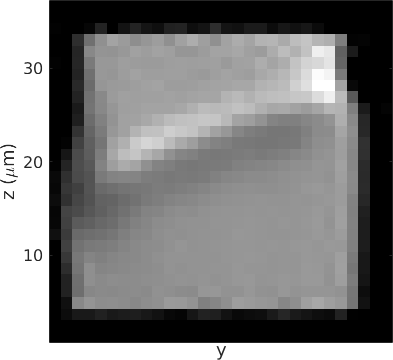}}
    \hspace{2pc}
    \subfigure[Computed "pseudo" volume]{\label{fig:3DTTrecvsdistortvoldistortvol}\includegraphics[width=15pc]{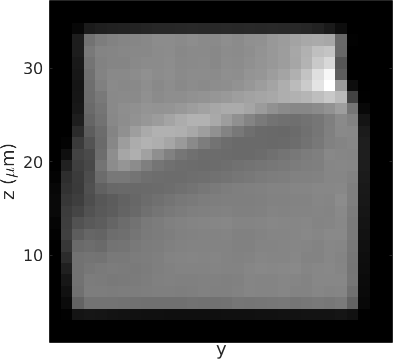}}
    \caption{The scattering intensities in the 2D slices of (a) the 3D reconstruction from the TT scan shown in figure~\ref{fig:phantomttspotdist50} and (b) the "pseudo" distorted phantom volume computed by~\eqref{eq:equigrainrec3D} based on the intragranular orientations.}
    \label{fig:3DTTrecvsdistortvol}
\end{figure}

As derived in  section~\ref{sec:3DTTrecdistortvol}, in the presence of intragranular orientation spread, the 3D reconstruction using integrated TT spots results in a distorted volume.
The TT scan with $D$=50 mm, as illustrated in figure~\ref{fig:phantomttspotdist50}, is used to perform a 3D reconstruction. The voxel size in the reconstruction is $\sim$ 1.22 $\mu$m, which is identical to the detector pixel size. Figure~\ref{fig:3DTTrecvsdistortvol3DTTrec} shows the reconstructed density (local scattering power) in a 2D volume slice of this 3D TT reconstruction. The reconstructed volume is distorted. Figure~\ref{fig:3DTTrecvsdistortvoldistortvol} shows the distorted phantom volume computed by~\eqref{eq:equigrainrec3D}, which is similar to the 3D TT reconstruction.

\subsection{Calculation of the orientation reconstruction error}

As for the 5D TT reconstruction, the 3D grain volume can be obtained by summing the 5D scattering intensity distribution function over the 2D space of the normal, as illustrated in~\eqref{eq:volumerecfrom6Drec}.
\begin{equation}
    \label{eq:volumerecfrom6Drec}
    X_{3D}^*(x,y,z)=\iiint_O X_{5D}^*(x,y,z,n_x,n_y)dn_xdn_y.
\end{equation}
Where $X_{3D}^*$ is the scattering intensity distribution function of the volume reconstruction. The 3D grain shape can be reconstructed by the 3D segmentation of $X_{3D}^*$.

Besides the shape, the average normal vector of each voxel can also be computed from $X_{5D}^*$. Let unit vector $\hat{\mathbf{n}}_{voxel}^*$=[$\bar{n}_x^*$, $\bar{n}_y^*$, $\bar{n}_z^*$]$^T$ represent the average normal vector for each voxel in the reconstruction. The first two elements of $\hat{\mathbf{n}}_{voxel}^*$ are calculated as~\eqref{eq:voxelaveorirecfrom6Drec}.
\begin{equation}
    \label{eq:voxelaveorirecfrom6Drec}
    \begin{bmatrix}
        \bar{n}_x^*\\
        \bar{n}_y^*\\
    \end{bmatrix}=
    \iiint_O \dfrac{X_{5D}^*(x,y,z,n_x,n_y)}{X_{3D}^*(x,y,z)}
    \begin{bmatrix}
    n_x \\
    n_y \\
    \end{bmatrix}
    dn_xdn_y.
\end{equation}

Let $\hat{\mathbf{n}}_{voxel}$ indicate the average normal vector for each voxel in the phantom. The reconstruction error of the average normal vector for each voxel is quantified by the angle between $\hat{\mathbf{n}}_{voxel}^*$ and $\hat{\mathbf{n}}_{voxel}$, which is used to assess the reconstruction quality in the following analysis.

\subsection{Using opposite scattering vectors}

Multiple TT scans can be acquired from an identical lattice plane, by using different diffraction vectors such as ($h\,k\,l$), ($\Bar{h}\,\Bar{k}\,\Bar{l}$) and ($2h\,2k\,2l$). Namely, the diffraction vector from a single plane can be changed by using the other side of the diffraction plane or using different diffraction orders or different beam energies. From an experimental point of view, since the sample goniometer tilt settings are identical for these different TT scans, only the detector height and base tilt angle need to be changed to perform these different scans. Since only one crystal plane is used, the joint reconstruction of these two TT scans can use a 2D orientation space of the normal vector, which is similar to a single TT scan reconstruction.

\begin{figure}[H]
    \centering
    \includegraphics[width=25pc]{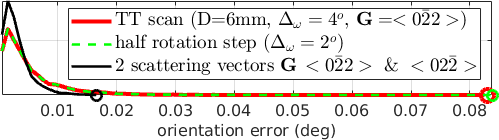}
    \caption{Histograms depicting errors of voxel average normal vectors $\hat{\mathbf{n}}_{p,voxel}^*$ between the phantom and the TT reconstructions.}
    \label{fig:hist2detsvs2xw}
\end{figure}

Figure~\ref{fig:hist2detsvs2xw} shows the histograms with a bin width of 0.001$^\circ$, showing the reconstruction errors of the average normal vectors for the voxels. The circles in figure~\ref{fig:hist2detsvs2xw} indicate the maximal errors. The three histograms correspond to the three reconstructions, which are two single-TT reconstructions with rotation intervals $\Delta_\omega$ of 4$^\circ$ and 2$^\circ$, respectively, and a joint reconstruction of 2 TT scans using two opposite scattering vectors with rotation interval of 4$^\circ$ ($\mathbf{G}$ $\langle0 \Bar{2} 2\rangle$ \& $\langle0 2 \Bar{2}\rangle$, $\Delta_\omega$=4$^\circ$). The data volume of a single TT scan conducted with a rotation interval of 2$^\circ$ is comparable to that of using two scattering vectors and a rotation interval of 4$^\circ$, since the halved rotation interval doubles the number of TT spots. The joint reconstruction of 2 TT scans using 2 opposite scattering vectors has fewer errors than the two single-TT reconstructions. The mean error in the average normal vectors of the voxels in the joint reconstruction is $\sim$ 0.003$^\circ$, while those of two single TT reconstructions are both $\sim$ 0.005$^\circ$. The base tilt sampling interval $\Delta_\phi$ is 0.03$^\circ$ and the detector distance $D$ is 6 mm. The reconstruction voxel size is the same as the pixel size of 1.22 $\mu m$, the orientation sampling resolution of the reconstructions is $\sim$ 0.009$^\circ$, and the number of the iterations is 100.

\begin{figure}[H]
    \centering
    \includegraphics[width=33pc]{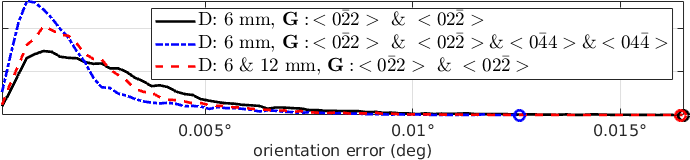}
    \caption{Histograms depicting errors in voxel average normal vectors $\hat{\mathbf{n}}_{p,voxel}^*$ between the phantom and the TT reconstructions ($\Delta_\phi$=0.03$^\circ$, $\Delta_o$=0.009$^\circ$). One reconstruction uses 2 detectors, and each of the other two reconstructions uses 4 detectors.}
    \label{fig:hist2detscomb}
\end{figure}

The joint reconstruction of more TT scans using the identical plane can be implemented by using the opposite scattering vectors, multiple diffraction orders, and multiple detector distances. Figure~\ref{fig:hist2detscomb} shows the histograms with a bin width of 0.0002$^\circ$, showing the errors of three joint reconstructions. The first joint reconstruction uses two TT scans using two opposite scattering vectors ($\mathbf{G}$ $\langle0 \Bar{2} 2\rangle$ \& $\langle0 2 \Bar{2}\rangle$); the second one uses four TT scans using the opposite scattering vectors and two diffraction orders ($\mathbf{G}$ $\langle0 \Bar{2} 2\rangle$ \& $\langle0 2 \Bar{2}\rangle$ \& $\langle0 \Bar{4} 4\rangle$ \& $\langle0 4 \Bar{4}\rangle$); the third one uses four TT scans using the opposite scattering vectors and two detector distances. Their mean orientation errors are about 0.003$^\circ$, 0.0019$^\circ$, and 0.0024$^\circ$. The more scattering vectors and more detector distances jointly used, the better the reconstruction, which is also explained in  section~\ref{sec:incrfftcomp}.

\subsection{Orientation sampling resolution}

\begin{figure}[H]
    \centering
    \includegraphics[width=36pc]{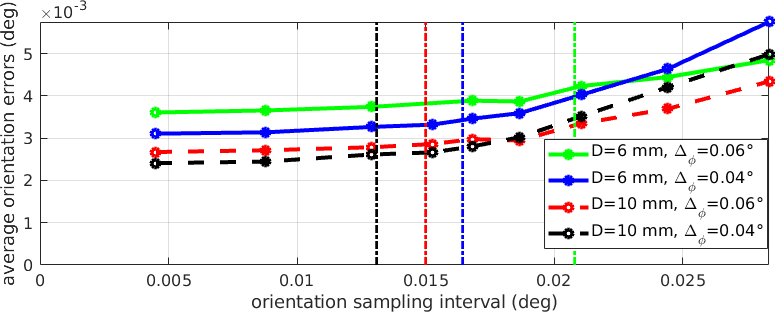}
    \caption{Four curves depict the relationship between orientation sampling interval $\Delta_o$ and the mean angular error of the average normal vectors for the voxels under four different scanning conditions. The vertical lines show the upper bounds of $\Delta_o$ calculated by~\eqref{eq:noaliasingoresmaxfreq}.}
    \label{fig:oriresorierrcurve}
\end{figure}

Figure~\ref{fig:oriresorierrcurve} shows the curves between orientation sampling resolution and mean errors in the average normal vectors for the voxels in the reconstructions. According to the curves, a finer orientation sampling resolution gives fewer errors. Moreover, if the orientation sampling resolution is sufficiently fine, a finer base tilt scan interval or a larger detector distance gives more precise orientation reconstructions. However, if the orientation sampling resolution is insufficient (e.g. due to computer memory limitations), the orientation reconstruction can be more inaccurate when the base tilt scan interval is finer or the detector is further out, as the curves have intersection points. As for the base tilt interval, it can be enlarged and adapted post-acquisition by interpolation in the 6D reconstruction algorithm.

The four scans corresponding to the four curves in figure~\ref{fig:oriresorierrcurve} are the joint 2-TT scans using two opposite scattering vectors ($\mathbf{G}$ $\langle0 \Bar{2} 2\rangle$ \& $\langle0 2 \Bar{2}\rangle$). The four scans use detector distances $D$ of 6 or 10 mm and base tilt scan intervals $\Delta_\phi$ of 0.04$^\circ$ or 0.06$^\circ$.~\eqref{eq:noaliasingoresmaxfreq} suggests a criterion that gives a upper bound for the orientation sampling interval. Here, the Bragg angle $\theta$ and the pixel size $\Delta_u$ are 5.687$^\circ$ and 1.22 $\mu m$. When the detector distance $D$ is 6 mm ($\Delta_u/C_p$=0.0577$^\circ$), the base tilt scan intervals $\Delta_\phi$ of 0.06$^\circ$ and 0.04$^\circ$ gives the upper bounds of 0.021$^\circ$ and 0.016$^\circ$; when the detector distance $D$ is 10 mm ($\Delta_u/C_p$=0.0346$^\circ$), the base tilt scan intervals $\Delta_\phi$ of 0.06$^\circ$ and 0.04$^\circ$ gives the upper bounds of 0.015$^\circ$ and 0.013$^\circ$. From the curves, we can see that, when the orientation sampling intervals are smaller than the suggested upper bounds, the orientation errors do not change much, and the curves have no intersection points.

\section{Discussion: Impact of approximations}
\label{sec:discussion}
In the experiments, the detector distance $D$, the Bragg angle $\theta$ and the intragranular orientation spread are usually less than 8 mm, 10$^\circ$ and 1$^\circ$. The detector pixel size can be 1.22 $\mu m$ - a typical value at the material science beamline ID11 at the European Synchrotron Radiation Facility (ESRF). The approximation in~\eqref{eq:smalletaondetapprox} results in an error in the projection position on the detector of approximately $(\Delta\eta^2 D\tan2\theta)/2<0.46 \mu m$, which is less than half the pixel size. The approximation in~\eqref{eq:diffdirapprox} results in an error in the diffracted beam direction of approximately $\Delta\eta\sin2\theta<0.35^\circ$. The voxel size used for grain volume reconstruction is usually the same as the detector pixel size. If the grain size is smaller than 50 $\mu m$, the diffracted beam direction error caused by the approximation in~\eqref{eq:diffdirapprox} leads to an error of less than $50\Delta\eta\sin2\theta$ $\mu m$ in the diffracted beam path within the grain volume, which is smaller than 1/4 voxel size. The approximation in~\eqref{eq:sam2diffcoordapproxphi} results in an error in the position of the grain voxel smaller than the product of grain size and intragranular orientation spread, which is smaller than the pixel size. The errors caused by the approximation are small, allowing the analysis of TT scan properties based on the approximated formulation. The smaller the grain size and the intragranular orientation spread, the closer the approximated formulation matches the real data.

A 5D reconstruction provides the 2D distribution of the normal vector for each voxel. For the study of polycrystals, typically, only the average normal vector for each voxel is required and the 5D reconstruction output is transformed into a vector field (average normal direction per voxel). Given the small voxel size, the orientation distribution function of each grain voxel usually has sparsity, resulting in sparsity in the pixel intensities of the 3D diffraction blobs recorded on the detector. Therefore, although a limited number of TT scans do not contain all the Fourier components to fully reconstruct the entire 5D space, it is still possible to achieve a meaningful reconstruction of the average normal vectors of the voxels.

In figure~\ref{fig:oriresorierrcurve}, when the orientation sampling resolution reaches the proposed lower bound, further increases in resolution only result in marginal improvements in reconstruction quality. The observed slight improvements occur because the derivation of this lower bound treats linear interpolation as a low-pass filter and approximates a cutoff frequency by disregarding the high-frequency tail of the spectrum.

Although not further discussed here, the combination of TT scans collected from different lattice planes (i.e. orthogonal directions)~\cite{liu2024characterization}, or the combination of DCT and TT acquisitions~\cite{vigano2020x} allows for reconstruction of the full (3D) lattice orientation using the 6D reconstruction framework and appropriate diffractometer coordinate transforms, accounting for the different sample tilt settings.



\section{Conclusions}

In this study, the TT projection geometry is theoretically analyzed to explore the limits of its reconstruction capacities and the methods to enhance them. Based on laminography and Radon transform, the mathematical expression for TT scans is derived, taking into account the intragranular lattice rotations and elastic strain in the direction of the scattering vector. Since a TT scan only uses a single crystal plane and scattering vector, one can only reconstruct the 3D grain shape, the local tilt of the plane normal and possibly the local lattice spacing (not demonstrated in this paper). When elastic strains are neglected, the grain is reconstructed in a 5D space, including 3 dimensions for the shape and 2 dimensions for the normal vectors.


Thanks to the simplified 6D TT scan expression derived in this article, and with proper integration of the TT spots along the base tilt scan, we demonstrated that the integrated TT spots can be interpreted as the projections of a "pseudo" grain volume distorted according to intragranular strain variations. This implies that the presence of structures within the orientation fields of the grain can also be detected and localized in the distorted volume of the 3D reconstruction from the TT scan.

Thanks to our Fourier analysis of the 5D TT scan expression, we show that a single TT scan contains only a subset of the Fourier transform components of the grain. That is, a single TT scan only contains partial information on the 5D/6D grain, which reduces the accuracy of the reconstruction. The joint use of multiple TT scans with varying detector distances and opposite scattering vectors can increase the Fourier transform components to improve the reconstruction. Using more TT scans extends the experiment time. We propose the joint use of two TT scans with opposite scattering vectors, leveraging the Friedel pairs of the TT spots, to better balance reconstruction quality with time efficiency. This approach is tested with simulated data, showing significant improvements in the reconstructions. Moreover, further improvements can be achieved by using more TT scans with varying detector distances or different scattering orders. As discussed in  section~\ref{sec:discussion}, this inadequate sampling of the Fourier space only marginally impacts our 5D TT reconstruction results, because the diffraction data have sparsity and we extract only one average orientation per voxel. If instead we wanted a complete 5D or even 6D TT reconstruction, we would necessarily need to switch to vertical line beam illumination. This in turn would require introducing a new scanning dimension (horizontal movement of the beam with respect to the sample), increasing the scan duration.

Based on the Fourier transform results, we also suggest a lower bound for the orientation sampling resolution in the 5D TT reconstruction.
This lower bound is validated using the simulated data, and it provides a practical guideline for choosing the correct reconstruction parameters, depending on the desired resolution.

In conclusion, this manuscript provides a fundamental understanding of TT acquisitions' capabilities for resolving local variations of lattice tilts in individual grains. The technique can be used to characterize the evolution of the orientation field during initial stages of plastic deformation. We provide to the users of this technique estimates regarding the accuracy of the reconstructions, and guidelines on the choice of reconstruction parameters.
Using the same mathematical framework, it will be possible to analyze similar X-ray diffraction based imaging techniques to better understand their capabilities and limitations.

\appendix

\section{Only lattice rotation}\label{app:TTformulaplast}

Ideally, in laboratory coordinates, the incident beam is along the $\hat{\mathbf{x}}_l$ direction and the base tilt axis is along the $\hat{\mathbf{y}}_l$ direction. When the Bragg condition is satisfied, the plane norm in laboratory coordinates can be calculated by~\eqref{eq:TTlabplnorm}~\cite{vigano2020x}. Ideally, when the base tilt angle is $\phi$=0, the rotation direction is the $\hat{\mathbf{z}}_l$ direction in the laboratory coordinates, so the coordinate [0,0,1]$^T$ is used here.
\begin{equation}
    \label{eq:TTlabplnorm}
    \hat{\mathbf{n}}_l=
    \begin{bmatrix}
        1 & 0 & 0 \\
        0 & \cos\Delta\eta & -\sin\Delta\eta \\
        0 & \sin\Delta\eta & \cos\Delta\eta \\
    \end{bmatrix}
    \begin{bmatrix}
        \cos\theta & 0 & -\sin\theta \\
        0 & 1 & 0 \\
        \sin\theta & 0 & \cos\theta \\
    \end{bmatrix}
    \begin{bmatrix}
        0 \\
        0 \\
        1 \\
    \end{bmatrix}=
    \begin{bmatrix}
        -\sin\theta \\
        -\sin\Delta\eta\cos\theta \\
        \cos\Delta\eta\cos\theta \\
    \end{bmatrix}.
\end{equation}
At the base tilt angle of $\phi$=$-\theta$, the rotation axis and the average normal of the plane are aligned along the $\hat{\mathbf{z}}_f$ axis in the reference coordinates. Due to the spread of intragranular orientation, the local normal of the plane can be different from the $\hat{\mathbf{z}}_f$-direction. If the unit vector $\hat{\mathbf{n}}_r$ = [$n_x$, $n_y$, $n_z$] indicates a local normal of the plane in the reconstruction coordinates, this normal in the laboratory coordinates can be calculated by~\eqref{eq:TTlocalplnorm2lab}.
\begin{equation}
    \label{eq:TTlocalplnorm2lab}
    \hat{\mathbf{n}}_l=\mathbf{\Omega}_\phi\mathbf{\Omega}_\omega\hat{\mathbf{n}}_r=
    \begin{bmatrix}
        \cos\phi & 0 & \sin\phi \\
        0 & 1 & 0 \\
        -\sin\phi & 0 & \cos\phi \\
    \end{bmatrix}
    \begin{bmatrix}
        \cos\omega & -\sin\omega & 0 \\
        \sin\omega & \cos\omega & 0 \\
        0 & 0 & 1 \\
    \end{bmatrix}
    \begin{bmatrix}
        n_x \\
        n_y \\
        n_z \\
    \end{bmatrix}.
\end{equation}
To fulfill Bragg's law,~\eqref{eq:TTlabplnorm} and~\eqref{eq:TTlocalplnorm2lab} need to be equal, which provides the diffraction condition expressed as~\eqref{eq:phietamatrixapprox}.
\begin{equation}
    \label{eq:phietamatrixapprox}
    \begin{bmatrix}
        -\sin\Delta\phi+2\sin^2(\Delta\eta/2)\sin\phi\cos\theta \\
        -\sin\Delta\eta\cos\theta \\
    \end{bmatrix}
    =\begin{bmatrix}
        \cos\omega & -\sin\omega \\
        \sin\omega & \cos\omega \\
    \end{bmatrix}
    \begin{bmatrix}
        n_x \\
        n_y \\
    \end{bmatrix}.
\end{equation}
Namely, from $\hat{\mathbf{n}}_l$=$\mathbf{\Omega}_\phi\mathbf{\Omega}_\omega\hat{\mathbf{n}}_r$, $\mathbf{\Omega}_\phi^T\hat{\mathbf{n}}_l$=$\mathbf{\Omega}_\omega\hat{\mathbf{n}}_r$ can be obtained. Since $n_x^2$+$n_y^2$+$n_z^2$=1, only two of these three equations are independent. Therefore, the equations are simplified as~\eqref{eq:phietamatrixapprox}, based on $\cos\Delta\eta$=$1-2\sin^2(\Delta\eta/2)$ and $\sin\phi\cos\theta$+$\cos\phi\sin\theta$=$\sin(\phi+\theta)$=$\sin\Delta\phi$.



By subtracting the expression on the left side of the equal sign from the expression on the right side in~\eqref{eq:phietamatrixapprox}, the vector function $\mathbf{f}_{\phi\eta}$ can be obtained as~\eqref{eq:phietamatrixapproxfunc}, so the constraint~\eqref{eq:phietamatrixapprox} is equivalent to the constraint $\mathbf{f}_{\phi\eta}$=$\mathbf{0}$.

From~\eqref{eq:phietamatrixapprox}, $\sin\Delta\eta$ can be expressed as~\eqref{eq:TTsineta}.
\begin{equation}
    \label{eq:TTsineta}
    \sin\Delta\eta = -(n_x\sin\omega+n_y\cos\omega)/\cos\theta.
\end{equation}
Since the intragranular misorientation angle is assumed to be sufficiently small, $\Delta\phi$ and $\Delta\eta$ are also sufficiently small that $\Delta\phi$ can be approximated as~\eqref{eq:approxphiplast}, based on~\eqref{eq:phietamatrixapprox} and first-order approximations.
The direction of the diffracted beam in laboratory coordinates can be calculated as~\eqref{eq:TTdiffdirlab}~\cite{vigano2020x}.
\begin{equation}
    \label{eq:TTdiffdirlab}
    \hat{\mathbf{d}}_l=
    \begin{bmatrix}
        \cos2\theta & -\sin\Delta\eta\sin2\theta & \cos\Delta\eta\sin2\theta \\
    \end{bmatrix}^T.
\end{equation}
Then, the direction in the reference coordinates can be calculated by $\hat{\mathbf{d}}_f$=$(\mathbf{\Omega}_\phi|_{\phi=\theta})\hat{\mathbf{d}}_l$, which results in~\eqref{eq:TTdiffdirdiffnoderiv}. The form of $\mathbf{\Omega}_\phi$ is described in~\eqref{eq:TTlocalplnorm2lab}. $(\mathbf{\Omega}_\phi|_{\phi=\theta})$ represents the matrix in the same form with $\mathbf{\Omega}_\phi$ when $\phi$=$\theta$.

As the direction of diffracted beam is obtained, the position of the voxel projection on the detector can be calculated. We first calculate the projection position [$u_0$, $v_0$]$^T$ of the voxel at the origin of the reference coordinates. $D/\cos2\theta$ calculates the distance between the origins of the reference coordinates and the detector coordinates. According to the geometric relationship of the special case, the diffraction direction expressed by~\eqref{eq:TTdiffdirdiffnoderiv} is parallel to [$D\cos\theta/\cos2\theta$, $u_0$, $D\sin\theta/\cos2\theta-v_0$]$^T$ in the reference coordinates, so $u_0$ and $v_0$ can be expressed as~\eqref{eq:TTdiffu0v0noderiv}.

It is assumed that $\hat{\mathbf{u}}$=$\hat{\mathbf{y}}$ and $\hat{\mathbf{v}}$=$-\hat{\mathbf{z}}$. Since the voxel at the origin of the reference coordinates is projected to [$u_0$, $v_0$]$^T$, so [$u$, $v$]$^T$ on the detector can receive the diffracted beam from the point $\mathbf{r}_{0,f}$ expressed by~\eqref{eq:refposdiffcoord} in the reference coordinates.
According to~\eqref{eq:TTdiffu0v0noderiv}, $\mathbf{r}_{0,f}$ can be expressed as~\eqref{eq:TTrefposdetailappendix}.
\begin{equation}
\label{eq:TTrefposdetailappendix}
    \mathbf{r}_{0,f}=
    \begin{bmatrix}
        0 \\
        u+D\sin\Delta\eta\tan2\theta \\
        -v \\
    \end{bmatrix}
    +\dfrac{2D\sin^2(\Delta\eta/2)\sin2\theta}{d_{f,x}}
    \begin{bmatrix}
        0 \\
        \sin\Delta\eta\tan2\theta\sin\theta \\
        1 \\
    \end{bmatrix}.
\end{equation}
Since $\Delta\eta$ and $D\Delta\eta^2$ are very small, $\hat{\mathbf{d}}_f$ and $\mathbf{r}_{0,f}$ can be approximated as~\eqref{eq:TTdiffdirapproxappendix}.
\begin{equation}
    \label{eq:TTdiffdirapproxappendix}
    \hat{\mathbf{d}}_f\approx
    \begin{bmatrix}
        \cos\theta & 0 & \sin\theta \\
    \end{bmatrix}^T;\,
    \mathbf{r}_{0,f}\approx
    \begin{bmatrix}
        0 & u+D\sin\Delta\eta\tan2\theta & -v \\
    \end{bmatrix}^T.
\end{equation}
The reference coordinates $\mathbf{r}_f$ can be calculated from the reconstruction coordinates as $\mathbf{r}_f$=$\mathbf{\Omega}_{\Delta\phi}\mathbf{\Omega}_\omega\mathbf{r}_r$, which can also be written as~\eqref{eq:possam2diffform1}.
\begin{equation}
    \label{eq:possam2diffform1}
    \mathbf{r}_f=
    \begin{bmatrix}
        x\cos\omega-y\sin\omega \\
        x\sin\omega+y\cos\omega \\
        z \\
    \end{bmatrix}-
    \begin{bmatrix}
        (x\cos\omega-y\sin\omega)(1-\cos\Delta\phi)-z\sin\Delta\phi \\
        0 \\
        (x\cos\omega-y\sin\omega)\sin\Delta\phi+z(1-\cos\Delta\phi) \\
    \end{bmatrix}.
\end{equation}
Since $\Delta\phi$ is very small, $\Delta\phi$ can be ignored so that $\mathbf{r}_f$ can be approximated as~\eqref{eq:sam2diffcoordapproxphi}.

In  section~\ref{sec:formulaonlyrot}, the TT scan is expressed as~\eqref{eq:TTformula}. Applying~\eqref{eq:TTdiffdirapproxappendix},~\eqref{eq:sam2diffcoordapproxphi} and the approximation of $\Delta\phi$ in~\eqref{eq:approxphiplast} to~\eqref{eq:TTformula}, the expression of the TT scan can be approximated as~\eqref{eq:TTformulaapprox}.


\section{Considering elastic deformation}\label{app:TTformulaelast}

Similarly to~\eqref{eq:TTlabplnorm}, when the Bragg's law is fulfilled in the presence of elastic strain, the normal of the plane in the laboratory coordinates can be expressed as~\eqref{eq:TTlabplnormelastic}.
\begin{equation}
    \label{eq:TTlabplnormelastic}
    \hat{\mathbf{n}}_l=
    \begin{bmatrix}
        -\sin\theta_e & -\sin\Delta\eta\cos\theta_e & \cos\Delta\eta\cos\theta_e \\
    \end{bmatrix}^T.
\end{equation}
From~\eqref{eq:TTlocalplnorm2lab} and~\eqref{eq:TTlabplnormelastic},~\eqref{eq:phietamatrixapproxelastic} can be obtained. By subtracting the expression on the right side of the equal sign from the expression on the left side in~\eqref{eq:phietamatrixapproxelastic}, the vector function $\mathbf{f}_{\phi\eta\Delta\theta}$ can be obtained as~\eqref{eq:phietamatrixapproxelasticfunc}, so the constraint~\eqref{eq:phietamatrixapproxelastic} is equivalent to the constraint $\mathbf{f}_{\phi\eta\Delta\theta}$=$\mathbf{0}$.
\begin{equation}
    \label{eq:phietamatrixapproxelastic}
    \begin{bmatrix}
        \cos\omega & -\sin\omega \\
        \sin\omega & \cos\omega \\
    \end{bmatrix}
    \begin{bmatrix}
        n_x \\
        n_y \\
    \end{bmatrix}=
    \begin{bmatrix}
        -\sin(\Delta\phi+\Delta\theta)+(1-\cos\Delta\eta)\sin\phi\cos\theta_e \\
        -\sin\Delta\eta\cos\theta_e \\
    \end{bmatrix}.
\end{equation}
From~\eqref{eq:phietamatrixapproxelastic},~\eqref{eq:TTsinetaelastic} can be obtained.
\begin{equation}
    \label{eq:TTsinetaelastic}
    \sin\Delta\eta = -(n_x\sin\omega+n_y\cos\omega)/\cos\theta_e;\,\Delta\phi \approx n_y\sin\omega-n_x\cos\omega-\Delta\theta.
\end{equation}
Similarly to~\eqref{eq:TTdiffdirlab}, the diffracted beam direction in the laboratory coordinates can be expressed as~\eqref{eq:TTdiffdirlabelastic}.
\begin{equation}
    \label{eq:TTdiffdirlabelastic}
    \hat{\mathbf{d}}_l=
    \begin{bmatrix}
        \cos2\theta_e & -\sin\Delta\eta\sin2\theta_e & \cos\Delta\eta\sin2\theta_e \\
    \end{bmatrix}^T.
\end{equation}
The diffracted beam direction in the reference coordinates can be calculated by $\hat{\mathbf{d}}_f$=$(\mathbf{\Omega}_\phi|_{\phi=\theta})\hat{\mathbf{d}}_l$, resulting in~\eqref{eq:TTdiffdirdiffelastnoderiv}. 

According to~\eqref{eq:TTdiffdirdiffelastnoderiv}, the origin of the reference coordinates is projected onto the detector position [$u_0$, $v_0$] expressed as~\eqref{eq:TTdiffu0v0elasticnoderiv}.

So, the detector position $(u,v)$ corresponds to the projection of the voxel at $\mathbf{r}_{0,f}$ in the reference coordinates, which is expressed as~\eqref{eq:TTrefposdetailappendixelastic}.
\begin{equation}
\label{eq:TTrefposdetailappendixelastic}
    \mathbf{r}_{0,f}\approx
    \begin{bmatrix}
        0 & u+D\sin\Delta\eta\tan2\theta & -v-D\sin2\Delta\theta/(\cos\theta\cos2\theta) \\
    \end{bmatrix}^T.
\end{equation}
Therefore, the TT scan can be approximated as~\eqref{eq:TTformulaapproxelastic}.


\section{TT scan using vertical line Beam}\label{app:MoreTTScanMethods}

In the experiment, the size of the cross section of the X-ray beam is limited by a vertical slit and a horizontal slit. Reducing the width of the horizontal slit can produce a vertical line beam. In the laboratory coordinates, the incident X-ray beam is along the $\hat{\mathbf{x}}_l$ direction, and the vertical gap is along the $\hat{\mathbf{z}}_l$ direction. The relative position between the vertical line beam and the grain can be changed by moving the diffractometer along the $\hat{\mathbf{y}}_l$-axis. Assume that the reference coordinates introduced in  section~\ref{sec:simpgeo} move together with the diffractometer along the $\hat{\mathbf{y}}_l$-axis. Let $l_y$ denote the position of the vertical line beam on the $\hat{\mathbf{y}}_l$-axis in the reference coordinates. The vertical line beam results in a diffraction constraint in the TT scan, which is expressed as $y_f$=$l_y$. Therefore, the expression of the TT spot changes from~\eqref{eq:TTformulaapprox} to~\eqref{eq:TTformulaapproxverticallinebeamappendix}.
\begin{equation}
    \label{eq:TTformulaapproxverticallinebeamappendix}
    \begin{split}
        & B_{TT}(u, v, \omega, \Delta\phi, l_y) \approx \iiint_V\iint_O X_{5D}(x,y,z,n_x,n_y)\delta(x\sin\omega+y\cos\omega-l_y) \\
        & \delta(\Delta\phi+n_x\cos\omega-n_y\sin\omega)\delta(z+v+(y\sin\omega-x\cos\omega)\tan\theta) \\
        & \delta((x+C_{p}n_x)\sin\omega+(y+C_{p}n_y)\cos\omega-u)dn_xdn_ydxdydz. \\
    \end{split}
\end{equation}
The Fourier transform $\hat{B}_{TT}$ of the TT spot $B_{TT}$ is expressed as~\eqref{eq:TTformulafftverticallinebeamappendix}, which contains the 5D Fourier transform components of the 5D grain $X_{5D}$.
\begin{equation}
    \label{eq:TTformulafftverticallinebeamappendix}
    \begin{split}
        & \hat{B}_{TT}(\rho_u, \rho_v, \omega, \rho_{\Delta\phi},\rho_{l,y}) \approx \iint_O\iiint_V X_{5D}(x,y,z,n_x,n_y)\exp(j\rho_vz\\
        & -j((\rho_u+\rho_{l,y})\sin\omega+\rho_v\tan\theta\cos\omega)x-j((\rho_u+\rho_{l,y})\cos\omega-\rho_v\tan\theta\sin\omega)y\\
        & -j(C_p\rho_u\sin\omega-\rho_{\Delta\phi}\cos\omega)n_x-j(C_p\rho_u\cos\omega+\rho_{\Delta\phi}\sin\omega)n_y)dVdO. \\
    \end{split}
\end{equation}



\begin{acknowledgements}
We thank B. Joste and B. Devincre for their help providing the discrete dislocation dynamics simulation of a grain containing a slip band, which has been used for the generation of synthetic diffraction data used in this work. 
\end{acknowledgements}

\begin{funding}
This research was funded by the Agence Nationale de la Recherche (ANR), grant number ANR 19-CE42-0017-02. A CC BY license is applied to the AAM resulting from this submission, in accordance with  the open access conditions of the grant.
\end{funding}

\ConflictsOfInterest{No potential conflict of interest was reported by the authors.}

\DataAvailability{The synthetic projection data used for the reconstructions in this paper can be made available upon request to the corresponding author.
}

\bibliography{article}

\end{document}